\documentclass[journal]{IEEEtran}
\usepackage{cite}
\usepackage{multicol,lipsum}
\usepackage{amsmath,amssymb,amsfonts}
\usepackage{algorithmic}
\usepackage{graphicx}
\usepackage{textcomp}
\usepackage{xcolor}
\usepackage{booktabs}
 \usepackage{hyperref}
\usepackage{enumerate}
\usepackage{algorithm}
\usepackage{multirow}
\usepackage{standalone}
\usepackage{tikz}
\usetikzlibrary{positioning}
\usetikzlibrary{decorations.markings}
\usepackage{caption}
\usepackage{subcaption}
 \usepackage{makecell}

\begin{document}
%
\title{AFDM-SCMA: A Promising Waveform for Massive Connectivity over High Mobility Channels
 }
%
%
%

\author{Qu Luo,~\IEEEmembership{Graduate Student Member,~IEEE,}
Pei Xiao,~\IEEEmembership{Senior Member,~IEEE,}
 Zilong Liu,~\IEEEmembership{Senior Member,~IEEE,}\\
 Ziwei Wan,~\IEEEmembership{Graduate Student Member,~IEEE,}
 Thomos, Nikolaos,~\IEEEmembership{Senior Member,~IEEE,} \\ 
  Zhen Gao,~\IEEEmembership{ Member,~IEEE,}  
 and
Ziming He,~\IEEEmembership{Senior Member,~IEEE.}
 \thanks{
 
 Qu  Luo  and  Pei  Xiao  are  with the  5G \& 6G  Innovation Centre, University of Surrey, U. K. (email: \{q.u.luo,   p.xiao\}@surrey.ac.uk). \\
  Zilong   Liu  and  Thomos, Nikolaos are   with   the   School   of   Computer   Science   and   Electronics   Engineering,   University   of   Essex,   U. K. (email:  \{zilong.liu, nthomos\}@essex.ac.uk).\\
  Ziwei Wan and Zhen, Gao are with the School of Information and Electronics, Beijing Institute of Technology, Beijing 100081, China, and Ziwei Wan is also with  the 5G \& 6G  Innovation Centre, University of Surrey, Guildford GU2 7XH, U. K.  Zhen Gao is  also with (BIT) Zhuhai 519088, China, also with the MIIT Key Laboratory of Complex-Field Intelligent Sensing, BIT, Beijing 100081, China, also with the Advanced Technology Research Institute of BIT (Jinan), Jinan 250307, China, and also with the Yangtze Delta Region Academy, BIT (Jiaxing), Jiaxing 314019, China (E-mails: : \{gaozhen16, ziweiwan\}@bit.edu.cn).\\
 Ziming He is with the Samsung Cambridge Solution Centre, System
LSI, Samsung Electronics, CB4 0DS Cambridge, U.K. (e-mail: ziming.he@
samsung.com).}
}
\maketitle

\begin{abstract}

This paper studies the affine frequency division multiplexing (AFDM)-empowered 
sparse code multiple access (SCMA) system, referred to as AFDM-SCMA,  for supporting massive connectivity in high-mobility environments.     First, by placing the  sparse   codewords on the AFDM chirp subcarriers,    the input-output   (I/O) relation of AFDM-SCMA systems is   presented.  Next, we delve into the generalized receiver design, chirp rate selection, and error rate performance of the proposed AFDM-SCMA.   The proposed AFDM-SCMA is shown to provide   a general framework and subsume  the existing
OFDM-SCMA    as a special case.    Third, for efficient transceiver design,  we further propose a class of sparse codebooks for simplifying the I/O relation,   referred to as    I/O relation-inspired codebook design in this paper.      Building upon these codebooks,  we propose a novel iterative detection and decoding scheme with   linear minimum mean square error (LMMSE)   estimator  for both downlink and uplink channels based on   orthogonal approximate message  passing principles.  Our numerical results demonstrate the superiority of the proposed AFDM-SCMA systems  over   OFDM-SCMA systems in terms of the error rate performance.   We show that the proposed receiver can significantly enhance the error rate  performance while reducing the detection complexity.

\end{abstract}


%
\IEEEpeerreviewmaketitle

\section{Introduction}

\subsection{Background}

\IEEEPARstart{I}{n}    recent years, there is a  surging demand   
for wireless connectivity    under various high mobility scenarios, e.g.,   integrated  terrestrial and non-terrestrial networks, high-speed trains and vehicle-to-everything communications \cite{NoorV2X,zhou2023overview}.  In addition to the inter-symbol interference over multipath fading channels, the communications  in these high mobility scenarios could also suffer  from the well-known time-varying channels with large  Doppler \cite{OFDMDoppler}.  In a low-earth orbit (LEO) satellite network, for instance, one needs to deal with time-varying high Doppler shift ranging from $-48$ KHz to $+48$ kHz for LEO altitude at $600$ km above the earth and with carrier frequency of 2 GHz. The  corresponding wireless channels are also called doubly selective channels. In this case, the  widely deployed  orthogonal frequency division multiplexing (OFDM) systems may be ineffective and unviable due to the significant     inter-carrier interference (ICI) \cite{OFDMDoppler}.

Against this background,  affine  frequency division multiplexing (AFDM)  has emerged as an   appealing
solution for efficient and reliable communications over high-mobility channels \cite{BemaniAFDM}.  The core idea of AFDM is built upon the       discrete affine  Fourier transform (DAFT)  which  was first  introduced in \cite{ErsegheAFDM}.  Specifically, the proposed AFDM system   employs quadratic exponential waveforms  (i.e., chirp-modulated signals) as mutually orthogonal basis subcarriers (SCs).  By contrast, OFDM systems rely   on a set of orthogonal SCs whose phases grow linearly with their individual carrier frequency indexes.   By   appropriately tuning the chirp rate according to the Doppler profile of the channel,  AFDM permits a separable and quasi-static channel representation, thus achieving  to achieve   the full   diversity  over the  doubly
selective channels. 
Moreover, every  AFDM system  is modulated/demodulated  via  inverse DAFT (IDAFT)/DAFT, which is a generalization of many other important  transforms, such as    inverse discrete Fourier transform (IDFT)/DFT and inverse discrete Fresnel transform (IDFnT)/DFnT \cite{FrFtAFT}. 
Notable examples include   OFDM  and orthogonal chirp division multiplexing (OCDM) \cite{OuyangOCDM}, where     IDFT/DFT and  IDFnT/DFnT are utilized  as the  modulation/demodulation, respectively.   More importantly,   IDAFT/DAFT can be efficiently implemented  by leveraging     OFDM modulation/demodulation  with two one-tap filters \cite{BemaniAFDM}.   These advantages of   AFDM are attractive for its widespread applications in beyond-5G (B5G)   wireless communication systems \cite{yin2022design,wang2022towards,NiAFDMSens}.

On the other hand, with the widespread proliferation of Internet-of-Things across every corner of this globe, new multiple access techniques with significantly improved spectrum efficiency are desired \cite{NGMA1}.  Over the past decade, non-orthogonal multiple
access (NOMA) has attracted tremendous research attention due to its capability of supporting       massive connectivity \cite{NGMA1,QuError}.  By allowing overloaded transmission of multiple users, existing  NOMA techniques can be mainly     categorized into power-domain NOMA  \cite{QUWCNC} and code-domain NOMA (CD-NOMA) \cite{QuError}. Among many others, this paper is concerned with  a disruptive CD-NOMA scheme, called  sparse code multiple access (SCMA), which  can achieve maximum-likelihood decoding performance with low complexity. Efficient message passing algorithm (MPA) \cite{WeijieSCMA} is adopted at the SCMA receiver    to exploit the codebook sparsity    as well as the    multidimensional constellation shaping gain \cite{lei2022scma,CDNOMA, luo2023design}.  \textit{The main aim of this work, therefore, is to study the integration of  SCMA and AFDM   for  providing ultra-reliable massive  connectivity in high mobility channels.}

  \vspace{-0.5em}
\subsection{Related Works}
There has been a  rich body of  research works on SCMA,  such as the  peak-to-average power ratio reduction for OFDM-SCMA system \cite{ChenSCMAOFDM},     codebook design \cite{lei2022novel, chen2022near, SCMAE2E,LiSCMA,ZYscmaCB}, and advanced receiver design\cite{chai2022improved,WeijieMIMOSCMA}.  However, these works were mostly  conducted under the assumption of static or   low mobility scenarios. Very recently, NOMA transmissions  building upon   orthogonal  time frequency space (OTFS) have  been studied  \cite{YaoSCMAOTFS,WenOTFS_SCMA,DekaOTFS_SCMA,ge2023lowMIMO,JHOP}.   The basic idea of OTFS is to send the data over   the delay-Doppler (DD) domain whereby every data symbol can enjoy the time-frequency diversity \cite{YuanOTFS1}. 
     Although power-domain NOMA (PD-NOMA) has been integrated with OTFS in  \cite{DekaOTFS_SCMA,OTFSNOMAGeYao}, the resultant systems require delicate user grouping and power allocation. Compared to SCMA, these might be challenging in high mobility channels due to small channel coherence time. Moreover, as shown in \cite{QuError}, SCMA enjoys high signal space diversity and hence outperforms power-domain NOMA in terms of the error performance.  By integrating SCMA with OTFS, it was shown in   \cite{DekaOTFS_SCMA} that OTFS-SCMA  provides improved  error   performance than conventional OFDM-SCMA and   OTFS-aided PD-NOMA. 
     However,  their proposed     two-stage detector consisting of a   linear minimum mean square error (LMMSE) detector and MPA in downlink channels  prevents the exploitation of the full potential of OTFS-SCMA system \cite{WenOTFS_SCMA}. 
To further enhance  the error performance,   an iterative structure between the LMMSE and MPA based on the orthogonal approximate messaging passing    (OAMP)  principles was developed in \cite{WenOTFS_SCMA}, yet at  the price of a  high computational complexity.  \cite{YaoSCMAOTFS} studied an SCMA empowered  coordinated multi-point vehicle communication system, where multiple mobile users are allowed to share the same DD resources with OTFS.  However, as  OTFS  suffers from the  high complexity of implementation and  excessive pilot overhead due to its 2D structure, the OTFS aided SCMA systems also face these challenges.
 
The  study of AFDM is still at its  early stage.     The performance of AFDM for high frequency band communication impaired with CFO and phase noise was  evaluated in  \cite{BemaniISWCS}.   Recently,   \cite{yin2022design}  studied the  orthogonal resource allocation, channel estimation and multi-user access scheme for    the AFDM    with multiple-input multiple-output  in high mobility scenarios.   \cite{zhu2023design,tao2023affine} studied the index modulation empowered AFDM schemes for enhancing   system spectrum efficiency.  A generalized MPA-based detection scheme was developed  in   \cite{wu2023message} to exploit  the  channel sparsity in  AFDM systems.  However, the   MPA  complexity   grows exponentially as the number of multipaths increases. In \cite{savaux2023dft}, instead of using DAFT, a   DFT-based modulation and demodulation techniques for AFDM was introduced, making   AFDM   backward compliant with with the  OFDM systems.   Moreover,  AFDM was also explored  for  integrated  sensing and communications in \cite{wang2022towards,NiAFDMSens}.   
Moreover, AFDM enjoys high backward compatibility with OFDM due to its 1D modulation nature instead of  the 2D  linear transform    in OTFS. This leads to reduced   implementation complexity and  channel estimation overhead compared to  OTFS \cite{BemaniAFDM,yin2022design}.
 
  \vspace{-0.5em}
\subsection{Motivation and contributions}
While the above works have revealed the superiority of AFDM, how it can enable improved    massive connectivity  over high mobility channels remains largely open.    
  Our primary objective  is to    study a novel SCMA empowered AFDM system, referred to as the AFDM-SCMA.    Different from existing works \cite{YaoSCMAOTFS,WenOTFS_SCMA,DekaOTFS_SCMA}  that primarily address uncoded systems, our focus is on coded systems.  The design of  AFDM-SCMA system is challenging   due to the following reasons:  1) in AFDM-SCMA systems, as different SCMA users may experience different radio propagation  channels, the AFDM parameters, such as the chirp rate, should be carefully tuned to attain full diversity;  2) Albeit  there are numerous SCMA codebooks reported in the    literature, they  are not optimized for AFDM-SCMA systems;  and 3) the  MPA might not be applicable to AFDM-SCMA because  its      complexity   grows exponentially with the number of multipaths. 
  

 The main contributions of this work are    summarized as follows:

\begin{itemize}
\item We propose a novel AFDM-SCMA system for supporting massive connectivity in high mobility environments. The proposed approach begins with the  allocation of SCMA codewords to AFDM SCs. Then, the   system model   for both downlink and uplink  AFDM-SCMA systems are given,  followed by a  generalized multi-user detection design.

\item We then delve  into an analysis of  the bit error rate (BER) performance of the AFDM-SCMA systems.  In contrast to the BER analyses in the  existing OFDM-SCMA systems, where every   SC  is assumed to be independent and identically distributed, we address a more realistic scenario involving multipath fading channels.    In addition,   to attain large diversity gain, the system parameters, such as the chirp rate,  in AFDM-SCMA are also studied.

\item  We propose a class of sparse codebooks for  simplifying the input-output (i.e., I/O) relation in AFDM-SCMA systems, referred to as   I/O relation-inspired codebook design. Building upon these codebooks,  we introduce a novel iterative detection and decoding scheme for both downlink and uplink channels based on the  OAMP  principles. Notably, the proposed receiver eliminates the need for an MPA decoder, resulting in a substantial reduction in decoding complexity in comparison to  the two-stage receiver proposed in previous works \cite{WenOTFS_SCMA,DekaOTFS_SCMA}.
\item Our numerical results demonstrate  the superiority of the proposed AFDM-SCMA systems. The results show that:  1) the proposed  AFDM-SCMA significantly outperforms  OFDM-SCMA    over doubly
selective channels;  2)  the derived analytical    BER curves match well with the simulated ones;  3)   the proposed transceiver  design can significantly enhance  the BER performance with reduced  the decoding complexity;  and 4) the proposed  I/O codebook and OAMP-assisted receiver can also be employed for OTFS-SCMA systems for performance enhancement and low-complexity detection. 
\end{itemize}

  \vspace{-0.5em}
\subsection{Organization}

This paper is organized as follows: Section \ref{PreLi} describes the preliminary concepts about SCMA and AFDM.   The proposed AFDM-SCMA is presented in Section \ref{SCMA_AFDM},  where the comprehensive details on the signal model, receiver design, AFDM-SCMA parameters, and an in-depth analysis of the BER performance are provided.  Section \ref{CBDec} introduces the proposed I/O relation-inspired codebook design and the proposed  advanced iterative detection and decoding scheme. The simulation results are presented and analyzed
in Section \ref{Sim}. Section \ref{conclu}  concludes the paper.

  \vspace{-0.5em}
\subsection{Notation}

The $n$-dimensional complex, real and binary vector spaces are denoted as $\mathbb{C}^n$, $\mathbb{R}^n$ and $\mathbb{B}^n$, respectively.  Similarly, $\mathbb{C}^{k\times n}$, $\mathbb{R}^{k\times n}$ and $\mathbb{B}^{k\times n}$  denote the $(k\times n)$-dimensional complex, real and binary  matrix spaces, respectively. ${{\mathbf{I}}_{n}}$ denotes an $n \times n $-dimensional  identity matrix.   $\text{diag}(\mathbf{x})$ gives a diagonal matrix with the diagonal vector of $\mathbf{x}$. $(\cdot)^\mathcal T$  and $(\cdot)^\mathcal H$ denote the transpose  and the Hermitian transpose operation, respectively.

\begin{figure*}
         \centering
  \includegraphics[width=0.62 \textwidth]{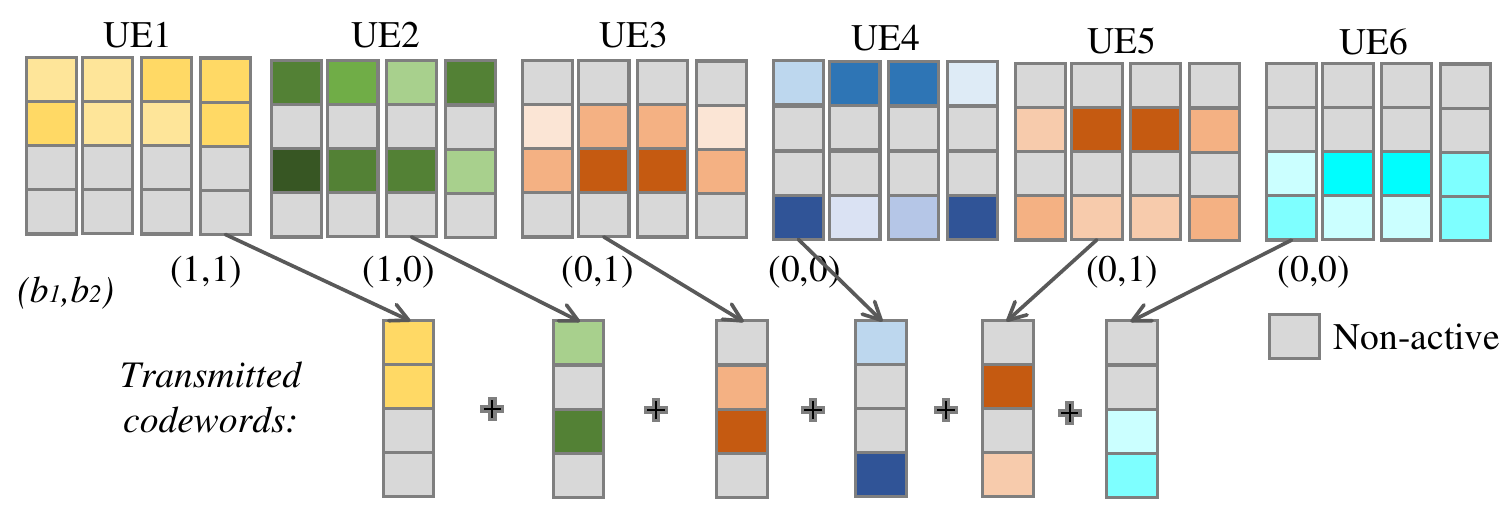}
         \caption{An example of SCMA encoding.}
         \label{SCMAMap}
           \vspace{-1em}
\end{figure*}

\section{PRELIMINARIES}
\label{PreLi}

\subsection{SCMA}

Consider a  $K \times J$  SCMA system, where $J $ users
communicate over $K$ resource nodes (RNs).  The overloading factor of SCMA system is defined by $\xi = J/K > 100 \%$, indicating  that the number of users that
concurrently communicate is larger than the total number of
orthogonal resources. 
Each user is  assigned   a unique codebook,  denoted by  $ \boldsymbol {\mathcal {X}} _{j}=\left [\mathbf{x}_{j, 1}, \mathbf{x}_{j,2}, \ldots, \mathbf{x}_{j,M}\right ]  \in \mathbb {C}^{K \times M}, j \in\{1,2, \ldots, J\}$, where $M$ denotes the codebook size and $\mathbf{x}_{j,m}$ is the $m$th codeword with   a dimension of $K$. The average codeword power of a codebook is assumed to be unit, i.e.,   $\text{Tr}(\boldsymbol {\mathcal {X}} _{j}^{\mathcal H} \boldsymbol {\mathcal {X}} _{j})=M$.  During the transmission, each user selects one codeword based on the input binary message. Denote  $\mathbf{b}_{j}=\left[b_{j, 1}, b_{j, 2}, \ldots, b_{j, \log _{2} M}\right]^{\mathcal {T}} \in \mathbb{B}^{\log _{2} M \times 1}$ by the input binary message  of the $j$th user. The SCMA encoding process of $j$th user can be expressed as \cite{SCMAE2E}
\begin{equation}
\small
\label{b2x}
    f_{j}: \mathbb{B}^{\log _{2} M \times 1} \rightarrow \boldsymbol{\mathcal { X }}_{j} \in \mathbb{C}^{K \times M}, \text { i.e., } \mathbf{x}_{j}=f_{j}\left(\mathbf{b}_{j}\right).
\end{equation}

\begin{figure}

         \centering
  \includegraphics[width=0.4 \textwidth]{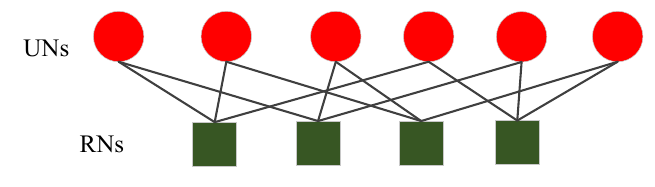}
         \caption{Factor graph representation of the SCMA system in (\ref{Factor_46}).}
         \label{facotGraph}
    \vspace{-1em}
\end{figure}

In SCMA,  the codebook sparsity is exploited by MPA for low-complexity decoding. Each $K$-dimensional complex codeword  has only $V$ non-zero elements, and the sparsity of the $J$ codebooks is reflected  by  the  sparse indicator   matrix   $\mathbf {F}_{K \times J}   \in \mathbb {B}^{K\times J}$.  An element of  ${\bf{F}}$ is defined as ${f_{k,j}}$ which takes the value of $1$ if and only if  the $k$th resource is occupied by  the $j$th user, and 0 otherwise.  A factor graph can graphically represents the sharing of the  resources among multiple  users.  Specifically, the factor graph of  a $K \times J$ SCMA system contains $J$ user nodes (UNs) and $K$  RNs. The $j$th UN is connected to the $k$th RN
  only if it occupies the $k$th resource. In addition, the number of users collides over a RN is denoted as $d_f$. In this paper, the following   indicator matrix   $J=6$, $K=4$, $d_f=3$ and $V=2$ is considered \cite{SCMAE2E}: 
  \begin{equation} 
 \label{Factor_46}
 \small
 {{\mathbf{F}}_{4\times 6}}=\left[ \begin{matrix}
   0  & 1 & 1 & 0 & 1 & 0  \\
   1 & 0 &1& 0 & 0 & 1\\
   0 & 1& 0 & 1 & 0 & 1 \\
   1 & 0 & 0 &1 & 1 & 0  \\
\end{matrix} \right],
  \end{equation}
where the  factor graph representation is shown in Fig. \ref{facotGraph}.

\subsubsection{Downlink channels} In downlink transmissions,  the codewords of $ J$ users are superimposed on $K$ RNs at the  base station (BS), i.e.,     \cite{SCMAE2E}
\begin{equation} 
\small
\label{x2w}
\mathbf {w} = \sum \limits ^{J}_{j=1}\mathbf {x}_{j},
\end{equation}
where $\mathbf {w} =[w_{1},w_{2}, {\dots },w_{K}]^{\mathcal T} \in \mathbb {C}^{K\times 1}$ is the superimposed codeword.  The received signal ${\mathbf {y}}_{u}=\left [{y_{u,1}, \ldots, y_{u,K}}\right]^{\mathcal T}$ at the $u$th user can be expressed as
\begin{equation} 
\small
\mathbf {y}_{u}=\text {diag}\left ({{\mathbf{h}}_{u}}\right)\mathbf {w}+\mathbf {n}_{u},
\end{equation}
where ${\mathbf{h}}_{u}=\left [{h_{u,1}, h_{u,2}, \ldots, h_{u,K}}\right]^{\mathcal T}$ is the channel gain between the BS and the $u$th user and  $\mathbf {n}_{u}$ is the Gaussian  noise vector  that has $\mathcal {CN}(0,N_0)$   entries, where $N_0$ is the noise variance.  For simplicity, the subscript $u$ is omitted throughout this paper. 

\subsubsection{Uplink channels} In contrast, for the uplink transmission, the $K \times 1$ received signal vector $\mathbf r$ at the BS can be written as
\begin{equation} 
\small
\mathbf r = \sum_{j=1}^{J} \text {diag}\left ({{\mathbf{g}}_{j}}\right)\mathbf {x}_j+\mathbf {n},
\end{equation}
where ${\mathbf{g}}_{j}  =\left [{g_{j,1}, g_{j,2}, \ldots, g_{j,K}}\right] ^{\mathcal T}$ denotes the channel vector from the $ j$th user and $\mathbf {n}$ the Gaussian  noise vector at the BS   which consists of $K$ entries each subject to independent  $\mathcal {CN}(0,N_0)$ distribution. 

 
\subsection{AFDM}
  This subsection first introduces   the digital implementation of AFDM and then presents the I/O relation.  For ease of presentation,    notations   $\mathbf x$, $\mathbf y$ and $\mathbf r$ are reused to describe the signal model of AFDM in this subsection. 






 \subsubsection{ AFDM modulation}  Denote $\mathbf x \in \mathcal C^{N \times 1}$ by the transmitted vector   in the DAFT domain, then the  modulated symbol is obtained by taking the  IDAFT  of $\mathbf x $, i.e., 
\begin{equation}
\small
\label{AFDMMOD}
    s[n] = \sum_{m =0}^{N-1} x[m] \varphi_{n}(m), n=0,1,\ldots,N-1,
\end{equation}
where $\varphi_{n}(m)$ is the AFT   kernel at the $m$-th SC and associated to the $n$-th modulated symbol. Specifically, we have 
\begin{equation}
\small
\varphi_{n}(m)={1\over \sqrt{N}}e^{j2\pi(c_{1}n^{2}+c_{2}m^{2}+{nm\over N})},
\end{equation}
where $ c_1 \geq 0$ and $ c_2 \geq 0$ are the AFDM parameters. Note that (\ref{AFDMMOD}) can be written in the matrix form as
 \begin{equation}
 \small
 \label{mod}
\mathbf s = \Lambda_{c_1}^{\mathcal H} \mathbf F^{\mathcal H} \Lambda_{c_2}^{\mathcal H}  \mathbf x = \mathbf A^{\mathcal H} \mathbf x,
\end{equation}
where $\mathbf A = \Lambda_{c_2}  \mathbf F \Lambda_{c_1}  $ is the DAFT matrix, $\Lambda_{c} =  \text{diag}\left ( e^{-j2\pi cn^2}, n=0,1, \ldots, N-1 \right)$, and $\mathbf F$ is the  DFT  matrix with entries $e^{-j2\pi mn/N} / \sqrt{N}$. 

In an AFDM system,   a prefix is required for combating multipath propagation and converting linear convolution into circulant convolution.  Different from   OFDM, in which  a cyclic prefix (CP) is utilized,   AFDM employs a chirp-periodic prefix (CPP) due to the inherent periodicity  of  DAFT.  The CPP with length of  $N_{\text{CPP}}$ is given by 
  \begin{equation}
  \small
    s[n] =   s[N+n] e^{-j 2 \pi c_1(N^2+2Nn)}, n=-N_{\text{CPP}}, \cdots,-1.
\end{equation}

 \subsubsection{Channel} We consider a doubly selective channel with the channel response at time $n$ and delay $l$ given by 
\begin{equation}
\small
\label{channel}
    g_{n}(l) = \sum_{p=1}^{P} h_p e^{-j\frac{2\pi}{N} \nu_p n} \delta(l-l_p),
\end{equation}
where $P$ is the number of paths, and $h_p$, $\nu_p$ and $l_p$ are the channel gain, Doppler shift  and the integer delay of the $p$th path. Note that $\nu_p$ is  normalized  with respect to the SC spacing and can be expressed as $\nu_p = \alpha_p +\beta_p$,   where $\alpha_p   \in [-\alpha_{\max},-\alpha_{\max}+1, \ldots,  \alpha_{\max}]$ and $\beta_p \in (-\frac{1}{2}, \frac{1}{2}]$    denotes the integer and fractional parts of $\nu_i$, respectively,  and $ \alpha_{\max}$ is the maximum integer Doppler.

  \subsubsection{AFDM demodulation}

At the receiver side, the received time domain signal can be expressed as 
\begin{equation}
\small
\label{rxSig}
    r[n] = \sum_{l=0}^{\infty} s[n-p]g_n[l] + v[n],
\end{equation}
where $v[n] \sim \mathcal{CN}(0, N_0)$  is the additive Gaussian noise. After discarding the CPP, (\ref{rxSig}) can be re-written in the matrix form as
\begin{equation}
\small
\label{rxSigM}
    \mathbf r  = \sum_{p=0}^{P} {\mathbf H_p} \mathbf s+ \mathbf v,
\end{equation}
where $\mathbf v $ is the noise vector in the time domain, ${\mathbf H_p} =  h_p \mathbf{\Gamma}_{\text{CPP}_p} \mathbf{\Delta}_{\nu_p} \mathbf \Pi^{l_p}$ is the  time domain channel matrix of the $p$th path, $\mathbf \Pi$ denotes the forward cyclic-shift matrix, i.e.,
\begin{equation} 
\small
\mathbf  \Pi = {\left[ {\begin{array}{ cccc} 0& \cdots &0&1 \\ 1& \cdots &0&0 \\ \vdots & \ddots & \ddots & \vdots \\ 0& \cdots &1&0 \end{array}} \right]_{N \times N}},
\end{equation}
$\Delta_{\nu_p} = \text{diag}(e^{-j \frac{2\pi}{N} \nu_p  }, n=0,1,\ldots,N-1)$ models the Doppler effect, and the effective CPP matrix $\mathbf{\Gamma}_{\text{CPP}_p}$ takes the following expression
\begin{equation}
\small
\label{TCPP}
  \mathbf{\Gamma}_{\text{CPP}_p}    = \text{diag} \left( \left\{\begin{matrix}
  e^{-j2\pi c_1 (N^2-2N(l_p-n))}, &   n< l_p, \\ 
  1, & n \geq l_p,
 \end{matrix}\right. \right).
\end{equation}
Finally, the received signal in the DAFT domain is obtained by applying the DAFT transform, i.e.,
\begin{equation}
\small
\label{DAFT}
 \mathbf y =  \Lambda_{c_2}  \mathbf F  \Lambda_{c_2}   \mathbf r   = \mathbf A \mathbf r.
\end{equation}

 \subsubsection{ The I/O relation} Substituting (\ref{AFDMMOD}), (\ref{channel})  and  (\ref{rxSig}) into (\ref{DAFT}), one has the I/O relation in the time domain given by 
\begin{equation}
\small
\begin{aligned}
     y[n] = \frac{1}{N} \sum_{m = 0}^{N-1} \sum_{p = 1}^{P} & h_p  \eta(l_p,n,m)  \gamma(l_p,\nu_p,n,m)  x[m] +v[n],
\end{aligned}
\end{equation}
where  
 \begin{subequations}
 \small
 \begin{align}
&\eta(l_p,n,m)  = {e^{j\frac{{2\pi }}{N}\left( {N{c_1}l_i^2 - m{l_i} + N{c_2}\left( {{m^2} - {n^2}} \right)} \right)}}, \\
& \gamma(l_p,\nu_p,n,m) =\frac{e^{-j2\pi (n-m+\text{Ind}_p + \beta_p)}-1}{e^{\frac{-j2\pi}{N} (n-m+\text{Ind}_p + \beta_p)}-1}, 
\end{align}
 \end{subequations}
 where $\text{Ind}_p = (\alpha_p + 2N c_1 l_p)_N$. The  I/O relation  in the DAFT domain in the matrix form  is given by 
\begin{equation}
\small
\begin{aligned}{\mathbf{y}} 
\label{inOut}
 & = \sum\limits_{p = 1}^P {{h_i}} \underbrace{{{\mathbf{\Lambda }}_{{c_2}}}{\mathbf{F}}{{\mathbf{\Lambda }}_{{c_1}}}{{\mathbf{\Gamma }}_{{\mathbf{CP}}{{\mathbf{P}}_p}}}{{\mathbf{\Delta }}_{{f_p}}}{{\mathbf{\Pi }}^{{l_p}}}{\mathbf{\Lambda }}_{{c_1}}^H{{\mathbf{F}}^H}{\mathbf{\Lambda }}_{{c_2}}^H}_{\mathbf H_p}  {\mathbf{x}} + {\mathbf{\tilde v}} \\ & = {{\mathbf{H}}_{{\text{eff }}}}   {\mathbf{x}} + {\mathbf{\tilde v}}, 
\end{aligned}
\end{equation}
where ${\mathbf{H}}_{{\text{eff }}} =\sum\nolimits_{p = 1}^P {{h_p}} \mathbf H_p $  is the effective channel matrix and ${\mathbf{\tilde v}}$ is the noise vector with the same distribution of ${\mathbf{ v}}$ as $\mathbf A$ is a unitary
matrix.  It can be shown that the  element of  $\mathbf H_p$ at row $n$ and column $m $ is 
 \begin{equation}
 \small
 \label{Hp}
 \begin{array}{l} {{\mathbf{H}}_p}[n,m] = \eta(l_p,n,m)  \gamma(l_p,\nu_p,n,m). \end{array} 
 \end{equation}
 
In AFDM systems, the parameters $c_1$ and $c_2$ can be adjusted so that the non-zero elements of matrix $\mathbf H_p$ in each path do not overlap within ${{\mathbf{H}}_{{\text{eff }}}}$, resulting in a comprehensive delay-Doppler channel representation. Further insights into the channel representation will be provided in the subsequent section  in conjunction with the SCMA structure.

\section{Proposed AFDM-SCMA}
 \label{SCMA_AFDM}
In this section, we present the proposed AFDM-SCMA framework for both downlink and uplink channels. We start by introducing the codeword allocation within AFDM-SCMA systems. Subsequently, we delve into the signal models and the design of multi-user detection for the proposed AFDM-SCMA schemes. Furthermore, the selection  of the AFDM-SCMA system parameters and  an in-depth analysis of the performance of AFDM-SCMA are also presented.

\subsection{SCMA Codewords Allocation }
\begin{figure}
    \centering
    \includegraphics[width=0.85\linewidth]{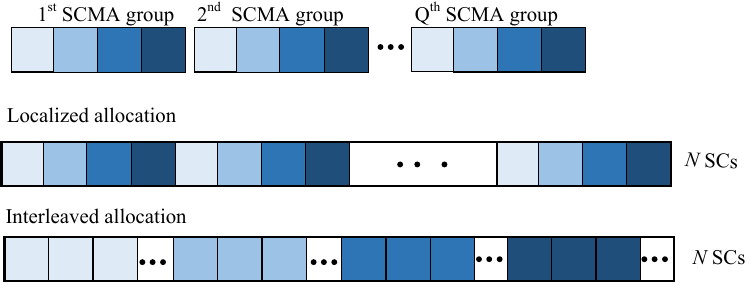}
    \caption{SCMA codewords allocation.}
    \label{SCMAallo}
      \vspace{-1em}
\end{figure}

In AFDM-SCMA,  multiple  SCMA groups with each group consisting of $K$ RNs and $J$ UNs  are transmitted over $N$ AFDM  SCs.    In this paper, we consider $Q$ SCMA groups that serve  $J$ users  and  assume that $N = QK$.  Denote   $\mathbf x_{j,q}$  by the $q$th SCMA codewords of user $ j$.  Let $ \bar{\mathbf x}_{\text{sym},j} $ consists of all the SCMA codewords to be allocated to the AFDM SCs, which takes the following expression:
\begin{equation}
\small
    \bar{\mathbf x}_{\text{sym},j} = [\mathbf x_{j,1}^{\mathcal T},\mathbf x_{j,2}^{\mathcal T},\ldots,\mathbf x_{j,Q}^{\mathcal T} ]^{\mathcal T}.
\end{equation} 
We further denote $  {\mathbf x}_{\text{sym},j}$ by the transmitted vector after the codewords allocation. To assign   the SCMA codewords to the AFDM SCs, we consider  the following two distinct  allocation schemes, as   depicted  in Fig. \ref{SCMAallo}, which discuss next.

\textit{a) Localized allocation:} The $N$ AFDM SCs are partitioned  into $Q=N/K$ groups, with each group consisting of $K$ SCs. Then, the $i$th SCMA codeword, i.e., $\mathbf x_{j,q}^{\mathcal T}, j=1,2,\ldots,J$ , is consecutively mapped to the $q$th AFDM group. 

\textit{b) Interleaved allocation:} The $N$ AFDM SCs are divided into $K$ groups, each consisting of    $Q$ SCs.  The $k$th entry of $\mathbf x_{j,q}$ is transmitted at the $q$th position of group $k$ over the corresponding  AFDM SC.


\subsection{Downlink AFDM-SCMA}
\label{DLSys}
\begin{figure}
    \centering
    \includegraphics[width=0.9\linewidth]{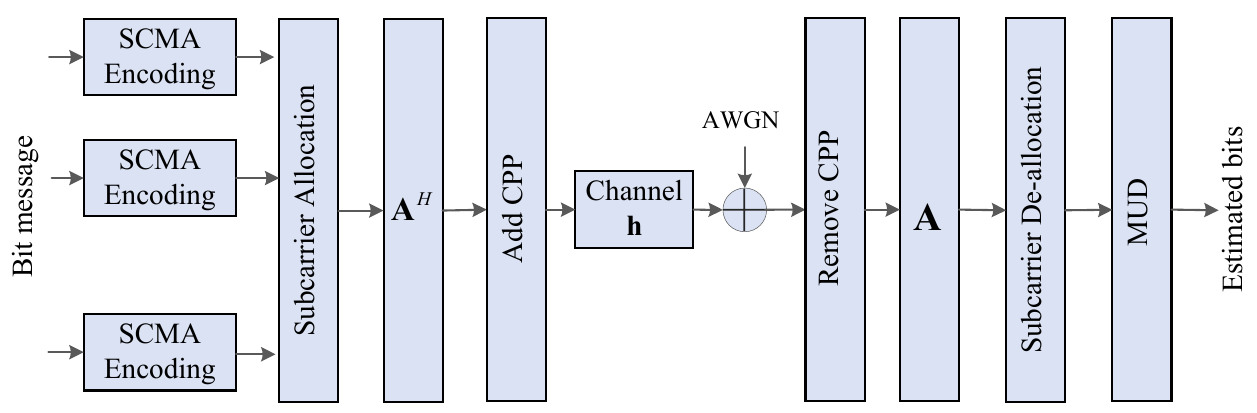}
    \caption{The proposed downlink AFDM-SCMA system.}
    \label{DLsystemmodel}
      \vspace{-1em}
\end{figure}

The proposed downlink AFDM-SCMA system is presented in Fig.  \ref{DLsystemmodel}.  The BS superimposes the symbol vectors of all users and performs codeword  allocations, leading to the superimposed codeword vector  as
\begin{equation}
\small
\label{Wsum}
    \mathbf w_{\text{sym}} =   \sum_{j=1}^{J} \mathbf x_{\text{sym},j}.
\end{equation} 
 Invoking the I/O relation of the AFDM system, the received signal  in the DAFT domain, denoted by  $ \mathbf y_{\text{sym}} \in \mathbb C^{N \times 1}$, can be written as 
\begin{equation}
\small
\label{YDown}
    \mathbf y_{\text{sym}} =   \mathbf H_{\text{sym}} \mathbf w_{\text{sym}}+\mathbf v_{\text{sym}},
\end{equation} 
where $\mathbf v_{\text{sym}} \in \mathbb C^{N \times 1}$ is the complex Gaussian noise vector with zero mean and variance of  $\mathbf I_N N_0$, and $\mathbf H_{\text{sym}} \in \mathbb C^{N\times N}$ denotes the complex channel coefficient matrix specifying the relationship defined in (\ref{inOut}). 

As evident from   (\ref{Wsum}) and (\ref{YDown}) that  $ \mathbf w_{\text{sym}}$ can be recovered from $\mathbf y_{\text{sym}}$ by applying  a simple LMMSE-based detector, i.e., 
\begin{equation}
\small
{\hat {\mathbf{w}}}_{\text {sum}}= \mathbf H_{\text{sym}}^{\mathcal H} \left [{  \mathbf H_{\text{sym}}  \mathbf H_{\text{sym}}^{\mathcal H}  +N_0^{2}{\mathbf{I}}_{N}}\right]^{-1} \mathbf y_{\text{sym}}.
\end{equation}
Subsequently, the estimated ${\hat {\mathbf{w}}}_{\text {sum}}$ will be fed to a standard MPA decoder to  obtain the detected  SCMA codewords $\hat{\mathbf x}_{\text{sym},j}, j=1,2,\ldots,J$.  

\textit{\textbf{Remark 1}: The above detection process is referred to as a two-stage detection scheme, similar to the one   in \cite{DekaOTFS_SCMA}.  It is worth noting that     such a  scheme    has  a considerably  high computational complexity.  In Section \ref{CBDec}, we will show that the detection can be simplified to   an one-stage LMMSE-based detector  by  joint design of    efficient sparse codebooks and advanced iterative 
 detection and decoding.}

\subsection{Uplink AFDM-SCMA}
\label{ULsec}
Consider an uplink AFDM-SCMA system with $J$ users communicate over $N$ AFDM SCs.    At the transmitter side, each user performs independent AFDM modulation with the same AFDM parameters $(c_1,c_2)$.  At the BS, the received signal after AFDM demodulation, denoted by $ \mathbf r_{\text{sym}} \in \mathbb C^{N\times 1}$, is given by
\begin{equation}
\small
\label{ULsig}
\begin{aligned}
       \mathbf r_{\text{sym}} & =   \sum_{j=1}^{J} \mathbf G_{\text{sym},j} \mathbf x_{\text{sym},j} +\mathbf n_{\text{sym}} \\
       &  = \mathbf G_{\text{all}} \mathbf x_{\text{all}} +\mathbf n_{\text{sym}},
\end{aligned}  
\end{equation} 
where $\mathbf G_{\text{sym},j} \in \mathbb C^{N \times N} $ denotes the effective channel matrix between the $j$th user and the BS, $\mathbf n_{\text{sym}} \in \mathbb C^{N\times 1}$ denotes the complex Gaussian noise vector with zero mean and variance of  $\mathbf I_N N_0$, and 
\begin{equation}
\small
\begin{aligned}
       {\mathbf{G}}_{\mathrm{all}} = &\left [{  \mathbf G_{\text{sym},1}, \mathbf G_{\text{sym},2}, \ldots, \mathbf G_{\text{sym},J} }\right] \in \mathbb C^{N \times NJ}, 
       \\ {\mathbf{x}_{\mathrm{all}}}  = &{\left [ \mathbf x_{\text{sym},1}^{\mathcal T},\mathbf x_{\text{sym},2}^{\mathcal T}, \ldots, \mathbf x_{\text{sym},J}^{\mathcal T} \right]^{\mathcal T}} \in \mathbb C^{NJ \times 1}.
\end{aligned}
\end{equation}
   Note that   $ \mathbf G_{\text{all}} $ and $\mathbf x_{\text{all}} $  are sparse vectors. Hence, a generalized MPA detector can be utilized for uplink AFDM-SCMA system.  Denote  $x_{\text{all},m}  $ by the $m$th codeword in $ \mathbf x_{\text{all}}$. Let 
   $\Omega_{\mathbf{G}_{\mathrm{all}}}{(n)} $ be the set of the columns at the  $n$th row of  $ \mathbf G_{\text{all}} $  connecting  to the transmitted codeword $x_{\text{all},m}$. In MPA, the major challenge is to compute the belief message from the observation nodes to the variable nodes.  In AFDM-SCMA systems, assume that the belief message  from the codeword node $m$  towards the observation node  $n$ is  denoted by $I_{c_{m} \rightarrow r_n}(x_{\text{all},m})$, then based on the sum-product rule, the message propagating from $r_n$ to $c_{m}$ is computed as follows \cite{luo2023design}:
\begin{equation}
  \small
  \label{FN_up}
 \begin{aligned}
 I_{r_n \rightarrow c_m} & (x_{\text{all},m})  =  \sum _{\substack{  x_{\text{all},m'} ,\\   m' \in   \Omega_{\mathbf{G}_{\mathrm{all}}}{(n)}  \backslash \lbrace m  \rbrace   }}  \frac{1}{\sqrt{2\pi N_0} } \\
& \text{exp} \left\{-\frac{1}{2N_0}  { \left \vert r_n -    \sum\limits_{m \in \Omega_{{\mathbf{G}}_{\mathrm{all}}(n)} }g_{n,m} x_{\text{all},m} \right \vert ^2} \right\}\\ 
& \quad \quad  \quad \quad   \times \prod _{ m' \in  \Omega_{\mathbf{G}_{\mathrm{all}}}{(n)}  \backslash \lbrace m  \rbrace }  I_{c_{m'} \rightarrow r_n}(x_{\text{all},m'}).
 \end{aligned}
\end{equation}
Due to the limited space, the update rule of $I_{c_{m'} \rightarrow r_n}(x_{\text{all},m'})$ is omitted in this paper. The readers are referred to \cite{MP} for more details about MPA.  The computational complexity  of MPA-based detector can be approximated as $\mathcal O(M^{\vert  \Omega_{\mathbf{G}_{\mathrm{all}}}{(n)} \vert})$.  

\textit{ \textbf{Remark 2:} The number of nonzero entries  at the $n$th row of  $ \mathbf G_{\text{all}} $, i.e., $\vert  \Omega_{\mathbf{G}_{\mathrm{all}}}{(n)} \vert$,  is determined by the  number of paths and the number of users that share the same  SCMA RN.  For integer Doppler, we have  $\vert  \Omega_{\mathbf{G}_{\mathrm{all}}}{(n)} \vert = d_f P$, where $d_f$ is the number of non-zero entries at each row of $\mathbf F$.  Obviously, the detection complexity increases exponentially with the increase  of $P$ and $d_f$. To this end, low complexity detection and decoding is further proposed in Section \ref{CBDec}. }


\subsection{AFDM-SCMA Parameters}

\begin{figure}[t]
\centering
\includegraphics[width=0.3\textwidth]{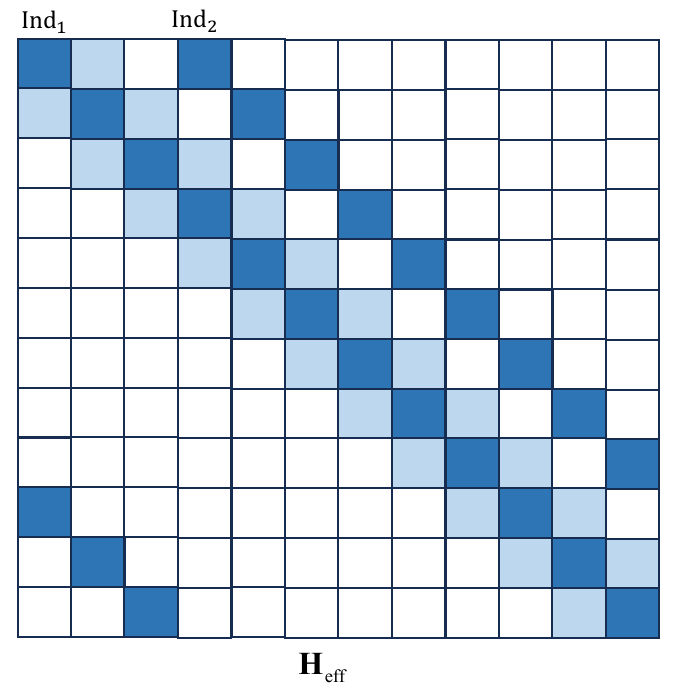}
\caption{Illustration of the effective channel matrix of a two-path channel  with the system parameters given by $K=4$, $N=12$, $l_1=0$, $l_2=1$, $\nu_1 = 0.05,  \nu_2 = 0 $ and $  k_{\nu } = 1 $.}
\label{SCMA_H}
\vspace{-1.5em}
\end{figure}

The performance of DAFT-based modulation schemes heavily depends  on the choice of parameters $c_1$ and $c_2$.  Recall (\ref{Hp}),  for each path, $\eta(l_p,n,m) $ has unit  energy,  and $\gamma(l_p,\nu_p,n,m)$ achieves the peak energy at $m = (n+\text{Ind}_p)_N$ and decreases as $m$ moves away from  $(n+\text{Ind}_p)_N$.  In this paper,  we   consider   $\gamma(l_p,\nu_p,n,m)$ is non-zero only for $m$ moves  $k_{\nu}$ away from  $(m+\text{Ind}_p)_N$.  Namely, the following holds
\begin{equation}
\small
\label{Hpath}
\begin{aligned}
   &  \vert {{\mathbf{H}}_p}[n,m] \vert   \\
   & = \left\{\begin{matrix}
  \vert \gamma(l_p,\nu_p,n,m)\vert    & \makecell { {{\left( {n + {{\operatorname{Ind} }_p}} - k_{\nu}\right)}_N} \leq m \\  \quad  \quad   \quad \leq {{\left( {n + {{\operatorname{Ind} }_p}} + k_{\nu}\right)}_N}} \\ 
 0 & {{\text{ otherwise }}}
\end{matrix}\right.
\end{aligned}.
\end{equation}
Fig. \ref{SCMA_H}  shows an example of    the effective  channel matrix of a downlink AFDM-SCMA system  over a two-path channel.  As evident  from (\ref{Hpath}), the location of each path depends on its delay-Doppler information and AFDM parameters.  It is essential to determine values for  $c_1$ and $c_2$ as such the  DAFT domain impulse response  forms a  full  delay-Doppler representation of the channel. In addition, $c_1$ and $c_2$ should also be adjusted to achieve the best diversity gain of AFDM-SCMA in doubly
selective channels.  Denote $l_p^{j}$ and $\nu_p^{j} = \alpha_p^{j} +\beta_p^{j}$ by the  delay and  the normalized Doppler of the $p$th path of the $j$th user, respectively,  where 
\begin{equation}
\small
\begin{aligned}
    \alpha_p^j  & \in [-\alpha_{\max},-\alpha_{\max}+1, \ldots,  \alpha_{\max}], \\
     \beta_p^{j} &  \in \Big(-\frac{1}{2}, \frac{1}{2} \Big], j=1,2,\ldots,J.
\end{aligned}
\end{equation}
 The number of paths for each user is denoted by $P_j$.  For the $j$th user, to achieve a large diversity gain,   the positions of  non-zero entries in the channel matrix for each path should not overlap with each other.  Therefore,  $c_1$ should satisfy \cite{BemaniAFDM}
\begin{equation}
\small
\label{c1eq}
     c_1  \geq  \max_{ 1 \leq j \leq J} \frac{{2(\alpha _{\max } + k_{\nu})}+1}{  2N \min_{p,q}(\vert l_p^j  - {l_q^j} \vert ) }.
\end{equation}

  It is shown in \cite{bemani2021afdm}  that the largest diversity of AFDM can be achieved for $c_1$ satisfying (\ref{c1eq}) and $c_2$ taken as   an arbitrary irrational number or a rational number sufficiently smaller than $ \frac{1}{2N}$.   The above analysis can be readily extended to uplink AFDM-SCMA systems. It is straightforward to show that (\ref{c1eq})  also holds for the  uplink AFDM-SCMA systems, respectively. 
 
\textit{\textbf{Remark 3:}  It is noted that the proposed  AFDM-SCMA can serve as a generic  framework and subsumes the existing OFDM-SCMA systems as special cases. Specifically, by setting $c_1 =0$ and $c_2=0$,  the  IDAFT and DAFT  boil down to the IFFT and DFT, respectively, and the proposed  AFDM-SCMA system is equivalent to the OFDM-SCMA system.}


\subsection{Performance Analysis}

 \begin{figure*}[t]
     \centering
 \begin{equation}
 \label{Hd}
\mathbf \Phi \left(  \boldsymbol{\Delta} \right) = \begin{bmatrix}
[H_1]_{1,\text{Ind}_1} \delta_{\text{Ind}_1} & & [H_P]_{1,\text{Ind}_P} \delta_{\text{Ind}_P} \\ 
 [H_1]_{1, (1+\text{Ind}_1)_N} \delta_{(1+\text{Ind}_1)_N}  & & [H_P]_{1, (1+\text{Ind}_P)_N} \delta_{(1+\text{Ind}_P)_N} \\ 
 \vdots &\cdots & \vdots\\ 
 [H_1]_{1, (N-1+\text{Ind}_1)_N} \delta_{(N-1+\text{Ind}_p)_N}& & [H_P]_{1, (N-1+\text{Ind}_P)_N} \delta_{(N-1+\text{Ind}_P)_N}
\end{bmatrix}.
\end{equation}
\vspace{-1em}
     \end{figure*}

In this section, we analyze the performance of the proposed AFDM-SCMA systems.  For simplicity, we start with the analysis in  the downlink AFDM-SCMA system  and then extend the results to uplink channels.  Recall  (\ref{YDown}), it  can be rewritten as 
\begin{equation}
\small
\label{DLR}
\begin{aligned}
    \mathbf y_{\text{sym}} & =    \sum_{p=1}^{P}  h_{p} \mathbf{H}_{p}  \mathbf w_{\text{sym}} +\mathbf v_{\text{sym}} \\
    & = \mathbf  \Phi(\mathbf w_{\text{sym}}) \mathbf h + \mathbf v_{\text{sym}},
\end{aligned}
\end{equation} 
where $\mathbf h= [h_1, h_2, \ldots, h_P    ]^{\mathcal T} \in \mathbb C ^{P \times 1 }$  specifies  the channel fading  gain of $P$ paths and $\mathbf \Phi(\mathbf w_{\text{sym}})$ is the $N  \times P$ concatenated matrix given by
\begin{equation}
\small
\label{phiW}
{\mathbf{\Phi }}(\mathbf w_{\text{sym}}) = \left[ {{{\mathbf{H}}_1}\mathbf w_{\text{sym}}| \ldots |{{\mathbf{H}}_P}\mathbf w_{\text{sym}}} \right].
\end{equation}
In (\ref{phiW}),  each row is  a cyclically shifted version of the first row depending on the Doppler tap values.
 Assume the  erroneously decoded codeword is  $\mathbf {\hat w}_{\text{sym}}$  when  $\mathbf w_{\text{sym}}$  is transmitted, where   $\mathbf {\hat w}_{\text{sym}} \neq \mathbf w_{\text{sym}}$.  Then,  the PEP between be two distinct data matrices $\mathbf {\hat w}_{\text{sym}}$  and  $\mathbf w_{\text{sym}}$   is given by 
\begin{equation}
\small
\label{PEP1}
\Pr\left( { \mathbf {\hat w}_{\text{sym}} \to \mathbf { w}_{\text{sym}} \mid {\mathbf{h}}} \right) = Q\left( {\sqrt {\frac{{{{\left\| {\mathbf \Phi \left(  \boldsymbol{\Delta} \right){\mathbf{h}}} \right\|}^2}}}{{2{N_0}}}} } \right),
\end{equation}
where $\boldsymbol{\Delta} = \mathbf {\hat w}_{\text{sym}} -\mathbf w_{\text{sym}} $,    ${\mathbf \Phi \left(  \boldsymbol{\Delta} \right)}$ is shown in (\ref{Hd})   and $\delta_i$ denotes the $i$th element of $\boldsymbol{\Delta}$.   By applying the approximation  $Q(x) \approx \frac{1}{{12}}\exp ({{ - {x^{2}}} / 2}) + \frac{1}{4}\exp ({{ - 2{x^{2}}} / 3})$ \cite{liu2021sparse},  we   can rewrite  (\ref{PEP1})  as
\begin{equation}
\small
\label{appro}
    \begin{aligned} \Pr\left( { \mathbf {\hat w}_{\text{sym}} \to \mathbf { w}_{\text{sym}} \mid {\mathbf{h}}} \right) 
 & \approx \frac{1}{{12}}\exp \left({ - \frac{  {{{{\left\| {\mathbf \Phi \left(  \boldsymbol{\Delta}   \right){\mathbf{h}}} \right\|}^2}}}}{4N_0}} \right) \\   & + \frac{1}{4}\exp \left({ - \frac{  {{\left\| {\mathbf \Phi \left(  \boldsymbol{\Delta}   \right){\mathbf{h}}} \right\|}^2}  }{3N_0}} \right).  \end{aligned} 
\end{equation}
Note that ${\mathbf \Phi \left(  \boldsymbol{\Delta}  \right)}^{\mathcal H} {\mathbf \Phi \left(  \boldsymbol{\Delta}  \right)}$ is a Hermitian matrix, its rank and the non-zero eigenvalues are
defined as $ R $ and $\lambda_i, i =1,2,\ldots, R$, respectively. Hence, it follows that 
\begin{equation}
\small
    \begin{aligned} {{\left\| {\mathbf \Phi \left(  \boldsymbol{\Delta}  \right){\mathbf{h}}} \right\|}^2} & = \mathbf h^{\mathcal H} \mathbf \Phi \left(  \boldsymbol{\Delta} \right)^{\mathcal H}\mathbf \Phi \left(  \boldsymbol{\Delta}  \right) {\mathbf{h}}  \\ 
    &   = {\mathbf{h}}^{H}{\mathbf{U\Sigma } }{{\mathbf{U}}^{H}}{{\mathbf{h}}} \\ & = {\mathbf{\tilde{h}}}^{\mathcal H}{\mathbf{\Sigma } }{{{\mathbf{\tilde{h}}}}_{j}} \\ & = \sum \limits _{i = 1}^{R} {{\lambda _{i}}{{\left| {\tilde{h}}_{ji} \right|}^{2}}},  \end{aligned}
\end{equation}
where $\mathbf U$ is a unitary matrix, $   \tilde{\mathbf h}  =   {\mathbf h}^{\mathcal H}\mathbf U $ and $\mathbf \Sigma = \text{diag} \{ \lambda_1, \lambda_2, \ldots, \lambda_R \} $.  Note that ${\tilde {\mathbf h}} $ has the same  distribution as ${ {\mathbf h}}$ since it is obtained by  multiplying a unitary matrix with ${ {\mathbf h}}$. We assume each path has independent Rayleigh distribution.  Hence,  by averaging (17) over the channel statistics, we arrive at
\begin{equation}
\small
\begin{aligned}
         \Pr  &  \left( { \mathbf {\hat w}_{\text{sym}} \to \mathbf { w}_{\text{sym}}} \right)     \approx \\  
         & \quad \quad  \quad \quad \frac{1}{{12}}\mathop \prod \limits _{i = 1}^{R} \frac{1}{{1 + \frac{{ {\lambda _{i}}}}{{4   N_0}}}} +  
         \frac{1}{4}\mathop \prod \limits _{i = 1}^{R} \frac{1}{{1 + \frac{{  {\lambda _{i}}}}{{3   N_0}  }}}.
\end{aligned}
\end{equation}
At high SNRs, i.e., $N_0 \to 0  $, it follows that 
   \begin{equation} 
 \small
 \label{pep3}
 \begin{aligned}
 \Pr     \left( { \mathbf w_{\text{sym}}^{m} \to \mathbf w_{\text{sym}}^{n}  } \right)    \approx N_{0}^{- R} \left ( \frac {4^{ R}}{12}  + \frac {3^{R }}{4} \right) \prod _{i=1 }^{R} {  \lambda_i^{-1}}.
 \end{aligned}
 \end{equation}
  One can see that    the rank of  $\mathbf \Phi \left(  \boldsymbol{\Delta} \right)$ determines the slope of the PEP curve. In general, $R$ is limited by the number of independent fading paths, which is also known as the diversity order \cite{luo2023enhancing}.  The   term  $\prod _{i=1 }^{R} {  \lambda_i }$ is  also known as the coding gain \cite{luo2023enhancing}.   The average BER can be formulated as \cite{liu2021sparse}
\begin{equation}
\small
\label{ABER}
\begin{aligned}
    &{P_{e}}\leq \\ & \frac{1}{{  M^{\frac{JN}{K}} \frac{JN}{K}{\log }_{2}}M   }    {\sum \limits _{ \mathbf  w_{\text{sym}}^{m} } {\sum \limits _{ \makecell{\mathbf  w_{\text{sym}}^{n}, \\\mathbf  w_{\text{sym}}^{m} \ne \mathbf  w_{\text{sym}}} }     \Pr  \left( { \mathbf w_{\text{sym}}^{m} \to \mathbf w_{\text{sym}}^{n}  } \right)   } }.   
\end{aligned}
\end{equation}
 where $\frac{1}{{  M^{\frac{JN}{K}} \frac{JN}{K}{\log }_{2}}M   } $ denotes the probability of each bit in  $\mathbf w_{\text{sym}}$. 
In the case of the uplink channels, assume the  erroneously decoded codeword is  $\mathbf {\hat x}_{\text{all}}$  when  $\mathbf x_{\text{all}}$  is transmitted, where   $\mathbf {\hat x}_{\text{all}} \neq \mathbf x_{\text{all}}$. To proceed, we rewrite (\ref{ULsig}) as 
\begin{equation}
\small
\begin{aligned}
    \mathbf r_{\text{sym}} & =  \sum_{j=1}^{J} \mathbf  \Phi(\mathbf x_{\text{sym},j}) \mathbf h_j + \mathbf v_{\text{sym}}\\
    & = \Omega(\mathbf x_{\text{all}}) \mathbf h_{\text{all}} + \mathbf v_{\text{sym}},
\end{aligned}
\end{equation}
where  $\mathbf \Phi(\mathbf x_{\text{sym},j})$ is the $N  \times P_j$ concatenated matrix defined in (\ref{phiW}), $\mathbf h_{\text{all}} = [\mathbf h_1^{\mathcal T}, \mathbf h_2^{\mathcal T}, \ldots, \mathbf h_J^{\mathcal T}]^{\mathcal T}$  and $ \Omega(\mathbf x_{\text{all}}) = [\mathbf \Phi(\mathbf x_{\text{sym},1}), \mathbf \Phi(\mathbf x_{\text{sym},2}), \ldots,\mathbf \Phi(\mathbf x_{\text{sym},J})] $.  We note that the signal  model in the uplink channels can be rewritten  in the form of (\ref{DLR}), hence, the average BER of the uplink AFDM-SCMA systems can be obtained analogue to (\ref{DLR})-(\ref{ABER}).

\section{Joint Codebook and Low-complexity Receiver Design}
 \label{CBDec}
This section investigates the transceiver design in AFDM-SCMA systems. Specifically, a class of efficient sparse codebooks is first presented to simplify the I/O relation of  AFDM-SCMA system. Subsequently, a low-complexity detection and decoding scheme is proposed by utilizing  the proposed codebooks.

\subsection{I/O Relation-Inspired Codebook Design}
We first introduce the codebook design in downlink AFDM-SCMA systems. Considering the signal model depicted in (\ref{Wsum}) and (\ref{YDown}), it is evident that a two-stage detector is essential at the  BS  to decode the transmitted codewords. This is due to the difficulties for integrating  the codebook superposition process, i.e.,  (\ref{Wsum}),  and the AFDM transmission process, i.e.,   (\ref{YDown}), into a   simplified signal model.  In the following, we address this problem through a novel codebook design.  

The main steps of the  proposed I/O relation-inspired codebook  design  can be summarized as follows:

 \textbf{Step 1: } Let us consider a  basic constellation alphabet set   $\mathcal {A}$. By drawing elements from $\mathcal{A}$,  denote its vector form  as $\mathbf a_{0} \in \mathbb C^{ M \times 1}$. 
 
 \textbf{Step 2: }   An initial $V$-dimensional  MC, denoted by $\mathbf A_{MC}$, is obtained by a   repetition  of  $\mathbf a _0$, i.e., $\boldsymbol {\mathbf A}_{MC} = \big[  \mathbf a _0^{\mathcal T}, \mathbf a _0^{\mathcal T}, \ldots, \mathbf a _0^{\mathcal T}\big]^{\mathcal T} \in \mathbb C^{V \times M}$.
 
 \textbf{Step 3: } Phase rotation  and power scaling are subsequently applied to each dimension of   $\boldsymbol {\mathbf A}_{MC}$ to generate multiple codebooks    for different users.  Specifically,  the $j$th user's codebook is generated by  
\begin{equation}
\small
     \boldsymbol{\mathcal X}_{j} = \mathbf {V}_{j} \mathbf {E}_{j} \mathbf{\Theta}_j {\mathbf A}_{MC},  \quad j=1,2,\ldots,J,
 \end{equation}
where  $\mathbf {V}_{j} \in \mathbb B^{K \times V}$  is the codebook mapping matrix that  maps a  $V$-dimensional dense constellation to  the $K$-dimensional sparse codebook $ \boldsymbol{\mathcal X}_{j}$, and    $\mathbf {E}_{j} = \text {diag} (  [ \sqrt{E_{j,1}},  \sqrt{E_{j,2}}, \ldots,  \sqrt{E_{j,V}}] ) \in \mathbb R^{V \times V}, 0 \leq \forall E_{j,v}\leq V,  \sum_{j=1}^{J}\sum_{v=1}^{V} E_{j,v} = J  $  
 and $ \mathbf{\Theta}_j =\text {diag} ([e^{i \theta _{j,1}},e^{i \theta _{j,2}}, \ldots,e^{i \theta _{j,V}}] ) \in \mathbb C^{V\times V},  0 \leq \forall \theta _{j,v} \leq 2\pi $   are  the power scaling and phase rotation matrices   of the $j$th user, respectively.   $\mathbf {E}_{j} $  and $\mathbf{\Theta}_j$  are designed to scale  and rotate    $\mathbf A_{MC}$  to enhance the constellation shaping gain, thereby improving overall codebook performance.    Based on the factor graph,  $\mathbf {V}_{j}$  can be constructed  according to the position of the zero  elements of  ${{\mathbf{f}}_{j}}$ by inserting  all-zero row vectors into the identity matrix ${{\mathbf{I}}_{V}}$. For example,  for the $\mathbf F$  given in ( \ref{Factor_46}), we have
\begin{equation}
\small
    {{\mathbf{V}}_{1}}=\left[ \begin{matrix}
  0 & 1  & 0 & 0  \\
  0 & 0 &  0 & 1  \\
\end{matrix} \right]^{\mathcal T} \text{and} \quad {{\mathbf{V}}_{2}}=\left[ \begin{matrix}
  1 & 0  & 0 & 0  \\
  0 & 0 &  1 & 0  \\
\end{matrix} \right]^{\mathcal T},
\end{equation}
  and $\mathbf {V}_{j}, j= 3,4,5$ and $6$ can be generated in a similar way.    
  
  Note that the phase rotation matrix and power scaling matrix can be combined together to form a constellation operator matrix, denoted as $\overline{\mathbf{Z}}_j =\mathbf{E}_j \mathbf{R}_j$.   We further combine the constellation operation matrix $\overline{\mathbf{Z}}_j$  and mapping matrix $\mathbf {V}_{j}$ together, i.e., $\mathbf{z}_{K \times 1}^j =\mathbf {V}_{j} \overline{\mathbf{Z}}_j \mathbf{I}_{K \times 1}$, where $\mathbf{I}_K$ denotes a column vector  of $K 1$'s. Hence,  the $J$ codebooks  can be represented by the signature matrix  $\mathbf{Z}= \left[ \mathbf{z}_{K \times 1}^1, \dots, \mathbf{z}_{K \times 1}^J \right]  $.   In this paper, the following signature matrix is employed:
   \begin{equation} 
   \small
 \label{signature_46}
 {\mathbf{Z}}=\left[ \begin{matrix}
   0 & z_1 & z _2 & 0 & z _3 & 0  \\
   z _1 & 0 & z _2 & 0 & 0 & z _3 \\
   0 & z _3& 0 & z _2 & 0 & z _1  \\
   z _3& 0 & 0 & z _2 & z _1 & 0  \\
\end{matrix} \right],
  \end{equation}
where $z_i = E_ie^{j \theta_i}, 1\leq i \leq d_f$   and $ \sum_{i=1}^{d_f} E_i = d_f/V$.  There are $d_f$  distinct rotation angles and and power scaling factors at  the each dimension of  (\ref{signature_46}) to distinguish the superimposed codewords.  With    (\ref{c1eq})   satisfied,  each SCMA codeword experiences the same multi-path channel as each path is well separated. Hence, the codebook design criteria becomes the minimum Euclidean distance (MED) of the superimposed constellation \cite{SCMAE2E,WenOTFS_SCMA}.  In this paper,  $E_i$ and $\theta_i,  1\leq i \leq d_f$ are optimized  to attain large MED.   Denote  ${\Phi }_{M^J} = \left\{\sum _{j=1}^{J}{{\mathbf{x}}_{j}} \vert \forall {{\mathbf{x}}_{j}}, 0 \leq j \leq J \right\}$ by the  superimposed  constellation,  then the MED of ${\Phi }_{M^J} $ is given by 
    \begin{equation}
 \label{dmin}
 \small
d_{\min }^2 = \min \left\{ {\parallel  \mathbf  w_n -  \mathbf {w}_m  \parallel}^2,   \forall {{\mathbf{w}}_{n}},{{\mathbf{w}}_{m}}\in {\Phi }_{M^J},   {{\mathbf{w}}_{n}} \neq {{\mathbf{w}}_{m}} \right\}.
  \end{equation}
Specifically, the codebook is optimized with the aid of the Matlab Global Optimization Toolbox. 

For uplink channels,  to simplify the I/O relation, we employ the low density signature.  The phase rotation in \textbf{ Step 3} is no long required as different users experience different channel conditions \cite{chen2022near}. Consequently, the codebook is generated by 
\begin{equation}
\small
     \boldsymbol{\mathcal X}_{j} = \mathbf {V}_{j}   {\mathbf A}_{MC},  \quad j=1,2,\ldots,J.
 \end{equation}
    The overall codebook design for downlink and uplink AFDM-SCMA   is summarized in \textbf{Algorithm 1}.

 \begin{algorithm}[t] 
\caption{I/O Relation-Inspired Codebook Design for AFDM-SCMA Systems}
\label{Adapt}
\begin{algorithmic}[1]
\REQUIRE {$V$, $\mathbf{F} $, ${{\mathbf{Z}}_{4\times 6}}$  and $\mathbf V_j, j= 1,2,\ldots, J$} \\
\STATE{ Choose a one-dimensional constellation $\mathbf a_0$,}
\STATE{ Obtain the $V$-dimensional MC  as $\boldsymbol {\mathbf A}_{MC} = \big[  \mathbf a _0^{\mathcal T}, \mathbf a _0^{\mathcal T}, \ldots, \mathbf a _0^{\mathcal T}\big]^{\mathcal T} \in \mathbb C^{V\times M}$,}
\STATE{\textbf{For uplink channels:}}
\STATE{ Generate the codebooks as $ \boldsymbol{\mathcal X}_{j} = \mathbf {V}_{j}   {\mathbf A}_{MC},  j=1,2,\ldots,J,$ }
\STATE{\textbf{For downlink channels:}}
\STATE{ Generate the codebooks as $  \boldsymbol{\mathcal X}_{j} = \mathbf {V}_{j} \mathbf {E}_{j} \mathbf{\Theta}_j {\mathbf A}_{MC}, j=1,2,\ldots,J,$ }
   \STATE{ Optimize   $\mathbf{Z} $  to attain large MED.  }
 \end{algorithmic}
   \vspace{-0.5em}
\end{algorithm}

   \subsection{Simplified I/O Relation for AFDM-SCMA}
   Based on the proposed codebooks, we present the simplified I/O relation for the proposed AFDM-SCMA systems. The constellation superposition process  in (\ref{x2w})   can be re-written as 
\begin{equation}
\small
\begin{aligned}
        \mathbf w  &= \sum\limits_{j=1}^J \mathbf x_j =\mathbf{Z}  \mathbf s,     
\end{aligned}
\end{equation}
   where $\mathbf s = [s_1,s_2, \ldots, s_J]^{\mathcal T}$, and $   s_j \in \mathcal {A}$  denotes transmitted symbol of the  $j$th user.  Namely,  the input   binary message of the  $j$th user are mapped to   symbol $s_j$ selected from a   constellation alphabet $\mathcal {A}$. Then the superimposed codeword $\mathbf w$ is obtained by multiplying the signature matrix with the input vector $\mathbf s$. The above signal model simplifies the I/O relation of SCMA encoding process   whilst maintaining a large MED.   Consequently, the   signal model of  (\ref{Wsum}) and (\ref{YDown}) can be re-written as 
  \begin{equation}
  \small
  \label{DLSig2}
  \begin{aligned}
    \mathbf y_{\text{sym},j}  & =    \mathbf H_{\text{sym},j}  \mathbf Z_{\text{sym}}  \mathbf s_{\text{sym}}+\mathbf z_{\text{sym},j}\\
    & = \overline {\mathbf H}_{\text{sym},j}     \mathbf s_{\text{sym}}+\mathbf z_{\text{sym},j},
  \end{aligned}
\end{equation} 
where  $\mathbf s_{\text{sym}} \in \mathbb C^{\frac{NJ}{K} \times 1}$ denotes the input data of $\frac{NJ}{K}$ SCMA groups, $\mathbf Z_{\text{sym}} \in \mathbb C^{N \times \frac{NJ}{K}}$ denotes the effective system signature matrix and $ \overline {\mathbf H}_{\text{sym},j} =  \mathbf H_{\text{sym},j}  \mathbf Z_{\text{sym}} \in \mathbb C^{N \times \frac{NJ}{K}} $ denotes the effective channel matrix for AFDM-SCMA.  For localized transmission,  $ \mathbf Z_{\text{sym}}$ is a block  diagonal  matrix, i.e.,  
   \begin{equation}
   \small
   \label{Zsys}
       \mathbf Z_{\text{sym}} = \text{blkdiag} \big \{ \underbrace {  \mathbf{Z},\mathbf{Z}, \ldots, \mathbf{Z}}_{N/K}\big \}. 
   \end{equation}

  Similarly, we can also obtain the  effective  signature matrix for the interleaved transmission with certain permutation of  $  \mathbf Z_{\text{sym}}$ in (\ref{Zsys}). 

\textit{\textbf{Remark 4:}  Based on   the proposed I/O relation-inspired codebook, the SCMA encoding process in (\ref{Wsum}) and (\ref{YDown})  is rewritten in the matrix form with an input dimension of  $ {\frac{NJ}{K}} $.   More importantly,  the   elements in $  \mathbf x_{\text{sym}}$ are independently drawn  from the same constellation $\mathcal A$ based on the incoming  bit messages, and  the proposed encoding process can still maintain a large MED.}

In uplink SCMA,   the   encoding process can be rewritten as 
\begin{equation}
\small
    \mathbf x_j  = \mathbf{f}_j  s_j,    \forall s_j \in \mathcal  {A}, j=1,2,\ldots, J,
\end{equation}
where $\mathbf f_j$ is the $j$th column of $\mathbf F$ defined in (\ref{Factor_46}). Hence, similar to the downlink channels, the signal model in (\ref{ULsig}) can be simplified to 
\begin{equation}
\small
\label{ULsig2}
\begin{aligned}
       \mathbf r_{\text{sym}} & =   \sum_{j=1}^{J} \mathbf G_{\text{sym},j} \mathbf F_{\text{sym},j} \mathbf s_{\text{sym},j} +\mathbf v_{\text{sym}}, \\
       &  = \overline{\mathbf G}_{\text{all}} { \mathbf s}_{\text{all}} +\mathbf v_{\text{sym}},
\end{aligned}  
\end{equation} 
where  $\mathbf s_{\text{sym},j} \in \mathbb C^{\frac{N}{K} \times 1}$ is the transmitted vector of the  $j$th user in an AFDM symbol,  $\mathbf F_{\text{sym},j}  \in \mathbb C^{N \times \frac{N}{K}}$  is the effective system signature matrix, $\overline{\mathbf G}_{\text{all}} =   \left [ \mathbf G_{\text{sym},1} \mathbf F_{\text{sym},1}, \mathbf G_{\text{sym},2} \mathbf F_{\text{sym},2}, \ldots,  \mathbf G_{\text{sym},J} \mathbf F_{\text{sym},J}  \right ] \in \mathbb C^{N \times \frac{NJ}{K}}$ is the simplified channel matrix of $J$ users, and  ${ \mathbf s}_{\text{all}} = \left[ \mathbf s_{\text{sym},1}^{\mathcal T}, \mathbf s_{\text{sym},2}^{\mathcal T}, \ldots,  \mathbf s_{\text{sym},J}^{\mathcal T} \right]^{\mathcal T} \in \mathbb C^{  \frac{NJ}{K}} \times 1$ collects input data of $J$ users.

\textit {\textbf{Remark 5:} Similar to that of the downlink channels, the input is a length $\frac{NJ}{K}$  vector, and the elements in ${ \mathbf s}_{\text{all}}$  are independently  mapped from the input binary messages and drawn from the same alphabet.  Compared to the signal model in (\ref{ULsig}),  (\ref{ULsig2}) reduces the dimensions of the channel matrix and  input vector from $ N \times NJ$ to $ {N \times \frac{NJ}{K}}$ and $NJ \times 1$ to $ \frac{NJ}{K} \times 1$, respectively.  Such a property allows    us to design advanced iterative receiver for the proposed AFDM-SCMA systems.}

    \vspace{-0.5em}
    \subsection{OAMP-assisted Iterative Receiver Design  for Coded AFDM-SCMA Systems}


By utilizing   the proposed I/O relation-inspired codebooks, we are able to simplify the I/O relations of AFDM-SCMA systems, as demonstrated in (\ref{DLSig2}) and (\ref{ULsig2}).  The simplified  signal models facilitate  the application of conventional linear receiver, such as  zero forcing and  LMMSE, for efficient symbol detection. In  this subsection, we introduce an iterative  receiver comprising an LMMSE estimator and an LDPC  decoder. Note that, the same scheme can be readily  extended to downlink channels.  The extrinsic information generated by the LMMSE estimator serves as \textit{a priori} information for the user-specific channel decoder, and conversely, the extrinsic information from the channel decoder is used as  \textit{a priori}  information for the LMMSE estimator.  

\begin{figure}[t]
\centering
\includegraphics[width=0.45\textwidth]{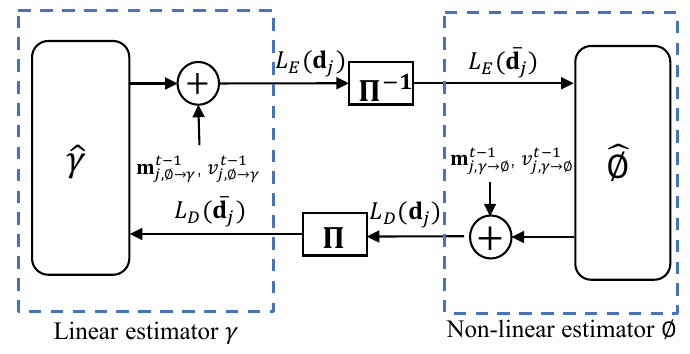}
\caption{Illustration of the  message propagation of  the $j$th user in the proposed OAMP-assisted iterative receiver.}
\label{FigItera}
\vspace{-1em}
\end{figure}

 It is important to note that both (\ref{ULsig2}) and (\ref{DLSig2}) represent overloaded signal models, due to its NOMA transmission nature.   To facilitate efficient message passing, new iterative receiver is designed by following the OAMP principles. This approach offers several advantages, including the capability of achieving the  LMMSE  capacity and the effectiveness in handling   ill-conditioned channel matrices and overloaded systems \cite{OAMP1}.   The key of OAMP is that the linear estimator   and the non-linear estimator    are designed to be orthogonal in terms of their individual estimation error vectors \cite{OAMP1}.  The orthogonality of LMMSE  and LDPC decoder avoids the estimated errors during  the iterative process. As a result, OAMP permits   the iterative process gradually converges  with stability.  For simplicity, denoted      $ \hat{\gamma}$ and $\hat{\phi}$ by the LMMSE detector and LDPC decoder, respectively. Similarly, denoted   $ {\gamma}$ and ${\phi}$ by the corresponding linear  and   nonlinear estimators, respectively. Fig. \ref{FigItera} shows the message propagation of user $j$ of the proposed OAMP-assisted iterative receiver. 

Denote  $\mathbf {d}_j = \left [ d^{1}_j,d^{2}_j, \ldots, d^{\log2(M)}_j \right]^{\mathcal T}$  by the  interleaved bits corresponding to the symbol
$s_j \in \mathcal {A}$ of user $j $ with $ j= \{1,2,\ldots,J\}$.     Note that   $\mathbf {d}_j$ is the interleaved coded bits from the channel encoder.  
We further denote  $P(\mathbf {d}_j)$ by the extrinsic probability  generated by  the LMMSE detector.  To initialize,   equiprobable   inputs are assumed without  \textit{a prior}  information from channel decoders. At each iteration, each  LMMSE detector    outputs   $P(\mathbf {d}_j)$ and demaps     $P(\mathbf {d}_j)$ to  the  bit-level  log-likelihood ratios (LLRs) as follows: 
\begin{equation}
\small
\label{LLR1}
\begin{aligned}
    L_{E}(d^{m}_j|  {s}_j) \triangleq  &\mathrm {ln}\frac {\sum _{\forall \mathbf {d}_j:d^{m}_j=0}  P(\mathbf {d}_j) } {\sum _{\forall \mathbf {d}_j:d^{m}_j=1}  P(\mathbf {d}_j)}. \\ 
\end{aligned}
\end{equation}
The resultant LLRs are further deinterleaved before  passing to the channel decoder.  Denote   $ L_{E} ({\overline {\mathbf {d}}_j})$  by the deinterleaved LLRs of the $j$th user. Upon receiving   $ L_{E} ({\overline {\mathbf {d}}_j})$, the  channel decoder computes  an estimation of the information bits  and outputs the  extrinsic LLRs. The output LLR of the  $i$th deinterleaved bit  is expressed as  \cite{OAMP1,OAMP2,OAMP3}
\begin{equation}
\small
\label{LLRLDPC1}
\begin{aligned}
    L_{D}({\bar {\mathbf {d}}_j[i]}) \triangleq \log \frac{\text{Pr}({\bar {\mathbf {d}}_j[i]} = 0 \vert L_{E} ({\bar {\mathbf {d}}_j})) }{\text{Pr}({\bar {\mathbf {d}}_j[i]} =1  \vert L_{E} ({\bar {\mathbf {d}}_j})) },
\end{aligned}
\end{equation}
where $ L_{D}({\bar {\mathbf {d}}_j})$ is the extrinsic LLR of the channel decoder output.  Similarly, the extrinsic LLRs are interleaved as $ L_{D}({ {\mathbf {d}}_j})$ and utilized as the input \textit{a priori}  information of the LMMSE detector.

\subsubsection{Messaging passing from LMMSE to channel decoder} 

 Following the LMMSE estimation principles,  the \textit{a posteriori} mean and covariance matrix at the $t$th iteration can be estimated as \cite{TuchlerMMSE}
\begin{equation}
\small
    \begin{aligned}
    \label{LMMSE1}
        \mathbf V_{\text{det}}^{t} & = \left( \overline{\mathbf G}_{\text{all}} \left(\mathbf V_{\phi \rightarrow \gamma}^{t-1 }\right)^{-1} \overline{\mathbf G}_{\text{all}} + \left(\mathbf V_{\phi \rightarrow \gamma}^{t-1}\right)^{-1}\right)^{-1}, \\
         \mathbf m_{\text{det}}^{t} &  =  \mathbf V_{\text{det}}^{t}  \left( \overline{\mathbf G}_{\text{all}}^{\mathcal T} \left( \mathbf V^{t-1}_{\phi \rightarrow \gamma } \right)^{-1}  \mathbf r_{\text{sym}}  +  \left( \mathbf V^{t-1}_{\phi \rightarrow \gamma } \right)^{-1}  \mathbf m_{\phi \rightarrow \gamma}^{t-1}\right),
    \end{aligned}
\end{equation}
where $\mathbf V^{t-1}_{\phi \rightarrow \gamma } = \text{diag  } \left ( [\mathbf v^{1,t-1}, \mathbf v^{2,t-1}, \ldots, \mathbf v^{Q,t-1} ] \right)$ and  $\mathbf m_{\phi \rightarrow \gamma}^{t-1} = [ \mathbf m_{\phi \rightarrow \gamma}^{1,t-1},  \mathbf m_{\phi \rightarrow \gamma}^{2,t-1},  \ldots, \mathbf m_{\phi \rightarrow \gamma}^{Q,t-1}   ]$  are    the     variance and mean matrices from the LDPC decoder  to the LMMSE detector at the  $t$-$1$th iteration, respectively, and   
$\mathbf m_{\phi \rightarrow \gamma}^{q,t-1} = [m_1^{q,t-1}, m_2^{q,t-1}, \ldots, m_J^{q,t-1}]$ and    $\mathbf v^{q,t-1} = [v_1^{q,t-1}, v_2^{q,t-1}, \ldots, v_J^{q,t-1}]$ are the corresponding  mean and variance  vectors at the $q$th blcok of user $j$.
For the  over-loaded signal model,  we utilize the matrix inverse lemma to reduce the computational complexity of  (\ref{LMMSE1}).  Specifically,  (\ref{LMMSE1}) can be rewritten as 
\begin{equation}
\small
    \label{LMMSE2}
    \begin{aligned}
        \mathbf V_{\text{det}}^t & = \mathbf V_{\phi \rightarrow \gamma }^{t-1} - \mathbf W_{\text{MMSE}}^t      \overline{\mathbf G}_{\text{all}}\mathbf V_{\phi \rightarrow \gamma}^{t-1},   \\
         \mathbf m_{\text{det}}^t &  =   \mathbf m_{\phi \rightarrow \gamma}^{t-1}  +  \mathbf W_{\text{MMSE}}^t \left(\mathbf r_{\text{sym}} -   \overline{\mathbf G}_{\text{all}}\mathbf m_{\phi \rightarrow \gamma}^{t-1}  \right),
    \end{aligned}
\end{equation}
where 
\begin{equation}
\small
    \mathbf W_{\text{MMSE}}^t = \mathbf V_{\phi \rightarrow \gamma}^{t-1}  \overline{\mathbf G}_{\text{all}}^{\mathcal T} \left( \mathbf V_{\phi \rightarrow \gamma}^{t-1}  +  \overline{\mathbf G}_{\text{all}} \mathbf V_{\phi \rightarrow \gamma}^{t-1}  \overline{\mathbf G}_{\text{all}}^{\mathcal T}  \right)^{-1}.
\end{equation}

Let $\mathbf v_{j,\text{det}}^{t} = [v_{j,\text{det}}^{1,t}, v_{j,\text{det}}^{2,t},\ldots,v_{j,\text{det}}^{Q,t}]^{\mathcal T}, \forall v_{j,\text{det}}^{q,t} \in  \mathbf V_{\text{det}}^t$ and   $\mathbf v_{j,\phi\rightarrow \gamma}^{t-1} = [v_{j,\phi\rightarrow \gamma}^{1,t-1}, v_{j,\phi\rightarrow \gamma}^{2,t-1},\ldots,v_{j,\phi\rightarrow \gamma}^{Q,t-1}]^{\mathcal T}, \forall v_{j,\phi\rightarrow \gamma}^{q,t-1} \in \mathbf V_{\phi \rightarrow \gamma}^{t-1}  $ be the  variance    vectors of the $j$th user at the $t$th and $(t-1)$-th iterations.  Let us define 
\begin{equation}
\small
\begin{aligned}
       \bar{v}_{j,\text{det}}^{t} & = \frac{1}{Q}  \Vert \mathbf v_{j,\phi\rightarrow \gamma}^{t}\Vert^2, \\ \bar{v}_{j,\phi\rightarrow \gamma}^{t-1} & = \frac{1}{Q} \Vert \mathbf v_{j,\phi\rightarrow \gamma}^{t-1}\Vert. 
\end{aligned}
\end{equation}Following the   message combining rule in OAMP \cite{OAMP1}, we  update the  mean and variance from the LMMSE to the LDPC decoder as 
\begin{equation}
\small
    \begin{aligned}
        {v}^t_{j, \gamma \rightarrow \phi} & = \left( \frac{1}{\bar{v}_{j,\text{det}}^{t}} - \frac{1}{\bar{v}_{j,\phi\rightarrow \gamma}^{t-1}} \right)^{-1},\\
       {\mathbf{m}}^t_{j,\gamma \rightarrow \phi} &=   {v}^t_{j, \gamma \rightarrow \phi} \left( \frac{ \mathbf m_{j,\text{det}}^t}{{\bar{v}_{j,\text{det}}^{t}}} - \frac{{\mathbf{m}}^{t-1}_{j,\phi\rightarrow \gamma}}{ \bar {v}^{t-1}_{j, \phi\rightarrow \gamma} } \right).
    \end{aligned}
\end{equation}
 To enhance algorithmic performance and regulate the convergence speed, we employ a damping factor $\kappa \in (0,  1]$ and update the mean and variance as \cite{MAMP}
 \begin{equation}
 \small
    \begin{aligned}
        {v}^t_{j, \gamma \rightarrow \phi} & = \left( \frac{\kappa }{\bar{v}_{j,\text{det}}^{t}} - \frac{1-\kappa }{\bar{v}_{j,\phi\rightarrow \gamma}^{t-1}} \right)^{-1},\\
       {\mathbf{m}}^t_{j,\gamma \rightarrow \phi} &=   {v}^t_{j, \gamma \rightarrow \phi} \left(\kappa  \frac{ \mathbf m_{j,\text{det}}^t}{{\bar{v}_{j,\text{det}}^{t}}} - (1-\kappa )\frac{{\mathbf{m}}^{t-1}_{j,\phi\rightarrow \gamma}}{ \bar {v}^{t-1}_{j, \phi\rightarrow \gamma} } \right).
    \end{aligned}
\end{equation}
Denote $s_{j}^{c}$ by the $c$th symbol in an LDPC block, then  based on the updated     mean  and variance, the  \textit{a posteriori} probability of    $s_{j}^{c}$ can be decomposed as follows  
\begin{equation}
\small
    P(s_{j}^{c} =  \mathcal C)   \varpropto \exp\left( -\frac{\vert \mathcal C - m_{j,\gamma \rightarrow \phi}^{c,t} \vert^2}{ {v}^t_{j, \gamma \rightarrow \phi} }   \right), \forall \mathcal C \in \mathcal {A}.
\end{equation}
 By (\ref{LLR1}),  the  extrinsic LLRs of the LMMSE can be obtained.  For QPSK constellation, the LLR  can be   computed efficiently as  \cite{TuchlerMMSE}
\begin{equation}
\small
\begin{aligned}
   L_{E}(d^{c,1}_j|  {s}^{c}_j)=&\frac {\sqrt {8}    \mathrm {Re}\{ m_{j,\gamma \rightarrow \phi}^{c,t}\}} { {v}^t_{j, \gamma \rightarrow \phi}   },  \\L_{E}(d^{c,2}_j|  {s}^{c}_j)=&\frac {\sqrt {8}   \mathrm {Im}\{m_{j,\gamma \rightarrow \phi}^{c,t}\}} { {v}^{t}_{j, \gamma \rightarrow \phi}  } .
\end{aligned}
\end{equation} 
 
\subsubsection{ Message passing from channel decoder to  LMMSE}  As mentioned, the  extrinsic LLR of LMMSE estimator is  deinterleaved before  passing to the channel decoder.   After channel decoding, the LDPC decoder outputs new extrinsic information $  L_{D}({\overline {\mathbf {d}}_j})$, defined in (\ref{LLRLDPC1}).  Through interleaver,  $  L_{D}({ {\mathbf {d}}_j})$  is obtained.   Upon obtaining the extrinsic LLRs from the channel decoder, $ L_{D}({ {\mathbf {d}}_j})$   is mapped to \textit{a prior} probabilities, i.e., 
\begin{equation} 
\small
\begin{aligned}
 P_D (s^c_j = \mathcal C) & = \prod_{i=1}^{\log_2(M)} \frac { \exp \left ( - \Psi_{\mathcal C}[i]  L_{D}({ {\mathbf {d}}_j}[c \log_2(M)+i])   \right)}{{1 + \exp \left (  L_{D}({ {\mathbf {d}}_j}[c\log_2(M)+i])   \right)}}\\
 & \varpropto \prod_{i=1}^{\log_2(M)}  \exp \left ( - \Psi_{\mathcal C}[i]  L_{D}({ {\mathbf {d}}_j}[c\log_2(M)+i])   \right),
 \end{aligned}
\end{equation}
where  $\Psi_{\mathcal C}[i] \in \{0,1\}$ denotes the labeling value of the $i$th bit of $\mathcal C$.   The  prior probabilities $P_D (\mathbf s_j )$ are then projected   into Gaussian distribution with its mean and variance respectively given by 
\begin{equation}
\small
\begin{aligned}
  {m}_{j, \text{dec}}^{c,t} \triangleq  & \sum_{\mathcal C\in{\cal{S}}}\, \mathcal C  P({s}_{j}^{c}= \mathcal C),\\
  v_{j, \text{dec}}^{c,t} \triangleq &   \sum_{\mathcal C\in{\cal{S}}}\, \vert \mathcal C\vert^2  P_D(s_j^c = \mathcal C)-\vert   {m}_{j}^{c} \vert^2.
\end{aligned}
\end{equation}
 For QPSK constellation, the mean and variance can be computed efficiently   by  as in \cite{TuchlerMMSE}
 \begin{equation}
 \small
 \begin{aligned}
    &   {m}_{j, \text{dec}}^{c,t}= \frac{\sqrt{2}}{2}\bigg(\tanh (L_{D}({ {\mathbf {d}}_j}[c\log_2(M)+1]) + \\
    & \quad \quad \quad \quad \quad \quad  i\tanh\left( L_{D}\left( \frac{{\mathbf {d}}_j [c\log_2(M)+2]}{2}\right)\right) \bigg),\\
     & {v}_{j, \text{dec}}^{c,t}= 1- \vert {m}_{j, \text{dec}}^{c}\vert ^2,
 \end{aligned}
 \end{equation}
  respectively. Let $ \mathbf { {m}}_{j, \text{dec}}^{t}$ and $ \mathbf { {v}}_{j, \text{dec}}^{t}$ denote the mean and variance vectors for each LDPC frame with its $c$th entry given by  $  {m}_{j, \text{dec}}^{c,t}$ and $ {v}_{j, \text{dec}}^{c,t}$, respectively. Define 
  \begin{equation}
  \small
\begin{aligned}
       \bar{v}_{j, \text{dec}}^{t} & = \frac{1}{B} \Vert \mathbf { {v}}_{j, \text{dec}}\Vert^2,
\end{aligned}
\end{equation}
 where $B$ is length of $\mathbf { {v}}_{j, \text{dec}}$, i.e. the number of symbols within an LDPC frame.  Then, the mean and variance are updated  respectively as
\begin{equation}
\small
\begin{aligned}
            {v}^t_{j, \phi\rightarrow \gamma}  &= \left( \frac{\kappa }{  \bar{v}_{j, \text{dec}}^{t}} - \frac{1-\kappa }{ {v}^{t-1}_{j, \gamma \rightarrow \phi} } \right)^{-1},\\
       \mathbf {m}^t_{j, \phi\rightarrow \gamma}  &=  {v}^t_{j, \phi\rightarrow \gamma}  \left(\kappa  \frac{ \mathbf { {m}}_{j, \text{dec}}^t}{\bar{v}_{j, \text{dec}}^{t}} - (1- \kappa )\frac{ {\mathbf{m}}^{t-1}_{j,\gamma \rightarrow \phi}}{  {v}^{t-1}_{j, \gamma \rightarrow \phi}} \right).
\end{aligned}
\end{equation}
 Finally, the new means and variances will feed back to the LMMSE estimator and utilized as \textit{a prior} information.

\subsection{State Evolution}

State evolution (SE) is a recursive procedure that tracks the mean square error (MSE) of local estimators during iterative processing.  In this subsection, we briefly present the SE in the proposed OAMP-assisted receiver design. The readers are referred to  \cite{OAMP1,OAMP2,OAMP3,LiuLeiOAMP2} for more details about the SE derivation in OAMP. To proceed, in what follows, we assume that the input and output errors of LE and NLE are independent of each other.  Let $ \tau^t$ and $\eta^t$ by the MSEs of  ${\mathbf{m}}^t_{\gamma \rightarrow \phi}$ and $ \mathbf {m}^t_{ \phi\rightarrow \gamma}  $, respectively. Namely, 
\begin{equation}
\small
   \tau_t= \frac{NJ}{K}\mathbb E \left [ \Vert{\mathbf{m}}^t_{\gamma \rightarrow \phi} - \mathbf s \Vert^2 \right ],
    \eta_t=\frac{NJ}{K}\mathbb E \left [ \Vert  \mathbf {m}^t_{\phi\rightarrow \gamma}  - \mathbf s \Vert^2 \right ].
\end{equation}
 
Then, SE refers to the following recursions of  $ \tau_t$ and $\eta_t$:
\begin{equation}
\small
\begin{aligned}
    \tau_t=  &  \gamma_{\text{SE}}(\eta_t) = \left( [\hat{\gamma}_{\text{SE}}(\eta_t)]^{-1} -\eta_t^{-1}\right)^{-1}, \\
  \eta_{t+1} = &  \phi_{\text{SE}}(\tau_t) =\left( [\hat{\phi_t}_{\text{SE}}(\tau_{t})]^{-1} -\tau_t^{-1}\right)^{-1},
\end{aligned}
\end{equation}
where 
\begin{equation}
\small
\begin{aligned}
     \hat{\gamma}_{\text{SE}}(\eta) & = \text{mmse}\{ \mathbf s \vert \sqrt{\eta}\mathbf s +\mathbf z_{\text{s}}, \gamma  \}, \\ 
     & =  \eta  - \frac{NJ}{K}\eta^2  \text{Tr}\left(  \overline{\mathbf G}_{\text{all}}^{H} \left(\eta \overline{\mathbf G}_{\text{all}} \overline{\mathbf G}_{\text{all}}^{H} +N_0\mathbf I \right)^{-1}\overline{\mathbf G}_{\text{all}} \right),  \\
\hat{\phi}_{\text{SE}}(\tau) & = \frac{1}{J}\sum_{j=1}^{J} \text{mmse}\{\mathbf s_j \vert  \mathbf s_j +\sqrt{\tau}\mathbf z_{\text{s}}, \phi \},
\end{aligned}
\end{equation}
where $\mathbf z_{\text{s}} \in \mathcal {N} (\mathbf 0, \mathbf I)$,  $\mathbf s_j$ is the modulated symbol for the $j$th NLE and $\text{mmse}\{a \vert b \} = \mathbb E \left\{ (a- \mathbb E (a \vert b) )^2\right \}$.  Unfortunately, it is quite difficult, if not impossible,  to obtain a close-form of  $\text{mmse}\{\mathbf s_j \vert  \mathbf s_j +\sqrt{\tau}\mathbf z_{\text{s}}, \phi \}$  for an NLE with LDPC codes. Here,   Monte Carlo simulation is employed to obtain the MSE of   NLE outputs.

\section{Numerical  Results}
 \label{Sim}
In this section, we present numerical simulation results  to validate the performance of  the proposed AFDM-SCMA systems.  The SCMA system with the indicator matrix given in (\ref{Factor_46}) is employed. 
The QPSK constellation is employed as the basic constellation, i.e., $\mathcal A = \{0.707+0.707i, 0.707-0.707i, -0.707+0.707i, -0.707-0.707i     \}$, for generating the I/O codebooks  and the resultant signature matrices  for uplink   and downlink channels are respectively given as $ {\mathbf{Z}}_{\text{UL}} = \mathbf F_{4 \times 6}$ and
   \begin{equation} 
   \small
 \label{signature_DL}
 {\mathbf{Z}}_{\text{DL}}=\left[ \begin{matrix}
   0 & 1.07i & 0.53 & 0 & 0.27 & 0  \\
    1.07i & 0 &0.53 & 0 & 0 & 0.27 \\
   0    & 0.27& 0 & 0.53 & 0 & 1.07i  \\
   0.27& 0 & 0 & 0.53 &  1.07i & 0  \\
\end{matrix} \right].
  \end{equation}

 \begin{table}  
\small
    \caption{Simulation Parameters}
    \centering
    \begin{tabular}{c|c}
    \hline
     \hline
       \textbf{ Parameters}  &  \textbf{ Values}  \\
        \hline
        \hline
        SCMA setting & $\xi = 150\%$ , $K=4, J=6$ \\
            \hline
        SCMA codewords allocation &  Localized and interleaved\\
          \hline
        Codebooks &  \makecell[c] {The proposed codebook \\ and Chen's codebook  \cite{chen2022near}}  \\ 
       \hline
       Channel model  &   \makecell[c] { EVA channel \cite{EVAc}} \\
         \hline
                 Number of AFDM SCs & $N = 128 $\\
                   \hline
                 CPP length &  $24$ \\
                  \hline
                Carrier frequency  & $4$ GHz\\
                 \hline
                 SC spacing  &  $15$  KHz\\
 \hline
                 User speed   & $300$  Km/h \\
 \hline
                 Maximum Doppler shift ($\nu_{\max}$) & 1.1 KHz \\
 \hline

    \end{tabular}
    \label{sim_para}
      \vspace{-1em}
 \end{table}

 \subsection{  Computational Complexity }
 
Since the same channel codes are employed for both the proposed OAMP-assisted receiver and the conventional turbo receiver,  we focus on comparing the computational complexities  of the multi-user detection modules.  The  computational complexity of the  LMMSE estimator is dominated by the matrix inverse in (\ref{LMMSE2}), which can be approximated as $\mathcal O (N^3)$.  The computational complexity of MPA for each SCMA group  is given by  $ \mathcal O (I_tKM^{d_f}d_f)$, where $I_t$ denotes the number of  MPA iterations.  Hence, the computational complexity of MPA for an AFDM-SCMA symbol is given by $ \mathcal O (I_tNM^{d_f}d_f)$.  In the proposed OAMP-assisted receiver, the MPA is no longer required, resulting in a complexity reduction ratio (CRR) approximated by
\begin{equation}
    \small
    \text{CRR} = 1- \frac{N^3}{I_tNM^{d_f}d_f+N^3}.
\end{equation}
For a system setting with   $N=64$, $d_f=3$ $I_t=6$ and $M=4$, the   CRR is given by $  \text{CRR} = 78\%$.  

   \vspace{-0.5em}
\subsection{Uncoded BER Performance}
\begin{figure} 
	\centering
	\begin{subfigure}{0.5\textwidth}
  \includegraphics[width=1.  \textwidth]{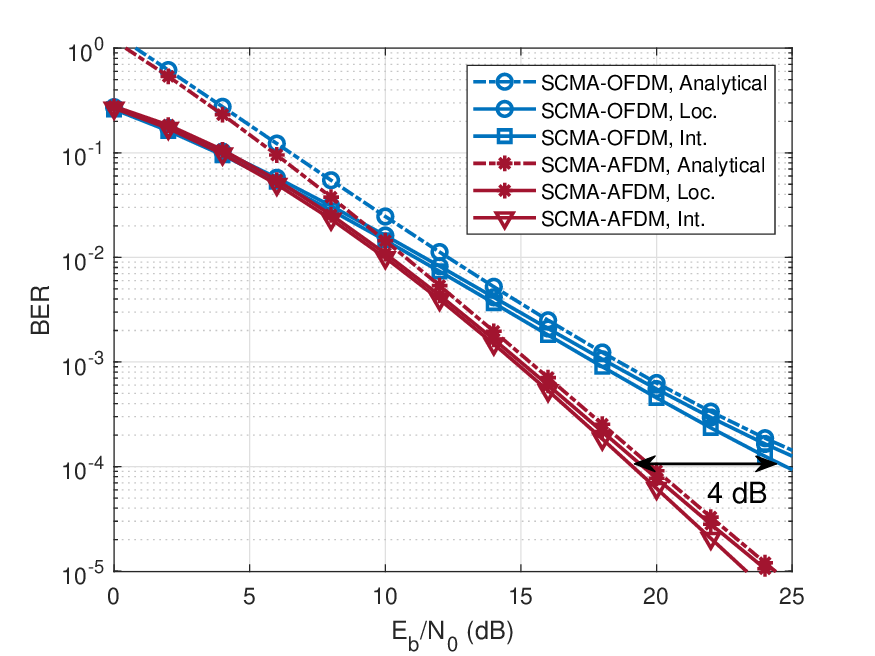}
		\caption{$ P=2$  }
	\end{subfigure}
	\begin{subfigure}{0.5\textwidth}
  \includegraphics[width= 1. \textwidth]{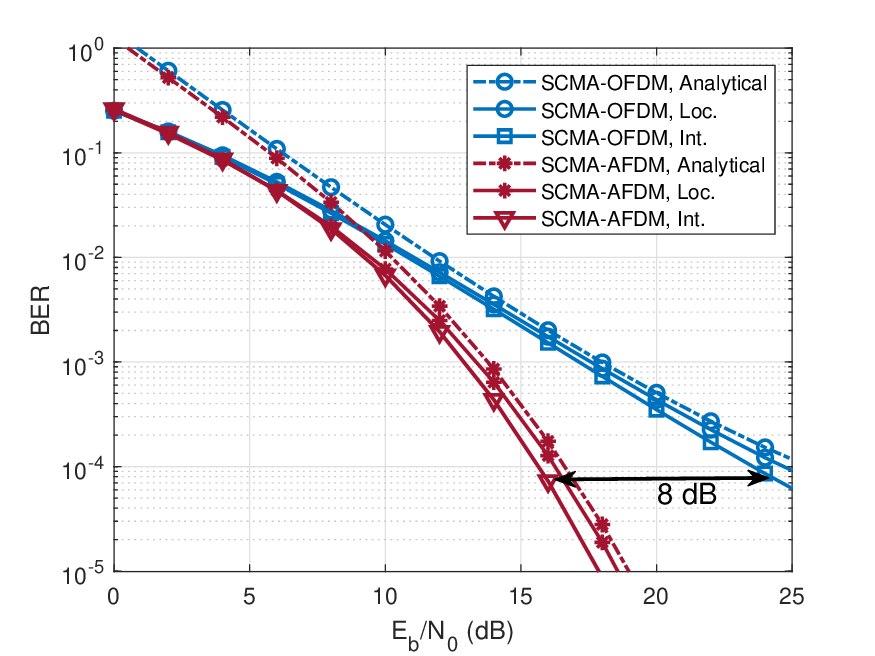}
		\caption{$P=3$}
	\end{subfigure}
	\caption{Simulated and analytical BER comparisons of the proposed AFDM-SCMA and OFDM-SCMA systems in uplink channels.}
	\label{SER_eq2}
 \vspace{-1em}
\end{figure}

We first present the simulated and analytical BER performance of the proposed  AFDM-SCMA in uplink channels with a generalized MPA detector  presented in  Section \ref{ULsec}.  Specifically, we  consider $J=6$ users  communicate over   $N=8$ AFDM SCs, and choose the  binary phase-shift keying (BPSK) as the basic alphabet $\mathcal {A}$.  Fig.  \ref{SER_eq2}   shows the BER performance of the proposed AFDM-SCMA and the conventional OFDM-SCMA systems over a two-path  and three-path  fading  channels, where each path is assumed to be independent and identically distributed with its variance given by $1/P$.   The curves with ``Int.'' and ``Loc.'' denote the interleaved and localized transmissions, respectively. ``CB'' denotes the abbreviation of codebook.       Notably, all the simulated BERs   match  well with the theoretical analysis at medium-to-high SNR values. At low   SNRs, a  discrepancy between the analytical and simulated BER is observed. This is due to the fact that the approximation in (\ref{appro}) is more accurate at high SNRs.     In addition,  the proposed AFDM-SCMA outperforms  conventional OFDM-SCMA significantly. Specifically, about $8$ dB $E_b/N_0$ gain   is observed for the   proposed AFDM-SCMA.

\begin{figure}[!t]
     \centering
   \includegraphics[width= 0.98\linewidth]{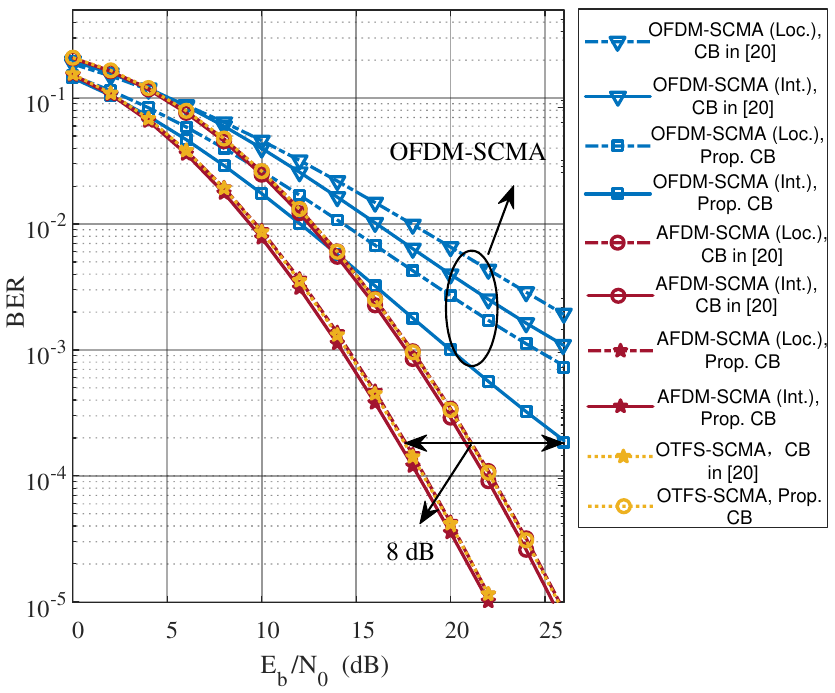}
  \caption{BER performance comparison of   AFDM-SCMA, OTFS-SCMA  and OFDM-SCMA   in downlink channels.}\label{DLBER}
  \vspace{-1em}
 \end{figure}

We now evaluate the performance of the  proposed  AFDM-SCMA in downlink channels with the two-stage detector  presented in  Section \ref{DLSys}.  The MPA iteration number is $5$,  and codebooks developed  in (\ref{signature_DL}) are employed. Refer to Table \ref{sim_para} for detailed simulation parameters. The EVA channel model is considered, where the power delay profile is given by $[0, -1.5, -1.4, -3.6, -0.6, $   $-9.1, -7.0, -12.0, -16.9]$  dB against the excess tap delays   $[0, 30, 150, 310, 370, 710, 1090, 1730, 2510]$ ns \cite{EVAc}.  The Doppler shift  at the $p$th   path of the $j$th user is  given by $\nu_{p,j} = \nu_{\max}\cos(\psi_{p,i})$ where $\psi_{p,i}$ has a uniform distribution $\psi_{p,i} \sim \mathcal U [-\pi, \pi]$. The OTFS-SCMA scheme proposed in \cite{DekaOTFS_SCMA} is also employed for comparison. Specifically, we consider the interleaved scheme with  the total numbers of the delay and the Doppler bins given by $8$ and $16$, respectively.  As seen from Fig.  \ref{DLBER},  the BER performance of OFDM-SCMA systems deteriorates  for the localized transmission scheme  compared to that of the   interleaved schemes due to the potential  deep fades within several consecutive  SCs. However,  the AFDM-SCMA with interleaved    and localized transmissions achieve similar BER performances.        This is due to 1) the intrinsic time-frequency spreading gain by AFDM transmission; 2) the chirp-rate tuning which leads to well separated multipath channels separated in the effective channel matrix. 
The proposed codebook achieves about $4$ dB gain over   Chen's codebook \cite{chen2022near} at the BER$=10^{-4}$ for  the proposed AFDM-SCMA system.  In addition, the proposed AFDM-SCMA achieves a   BER performance very close to that of  the OTFS-SCMA scheme as both can achieve the full channel diversity.   However, as mentioned earlier, the proposed AFDM-SCMA enjoys the benefits of  lower  implementation complexity and channel estimation overhead compared to OTFS-SCMA  \cite{BemaniAFDM,yin2022design} . In the next  subsection, we will   show that the proposed I/O codebook and low-complexity receiver can also be utilized for OTFS-SCMA systems.   Moreover, the proposed AFDM-SCMA outperforms the OFDM-SCMA system significantly. In particular,  $8$ dB gain is observed with the proposed codebook at the  BER$=2 \times 10^{-4}$.  Again, it is emphasized   that the   OFDM-SCMA system is the special case of the proposed systems with $c_1=0$ and $c_2=0$.

 \label{eqd}

\subsection{Coded BER Performance}

  \begin{figure}[!t]
     \centering
   \includegraphics[width= 0.95\linewidth]{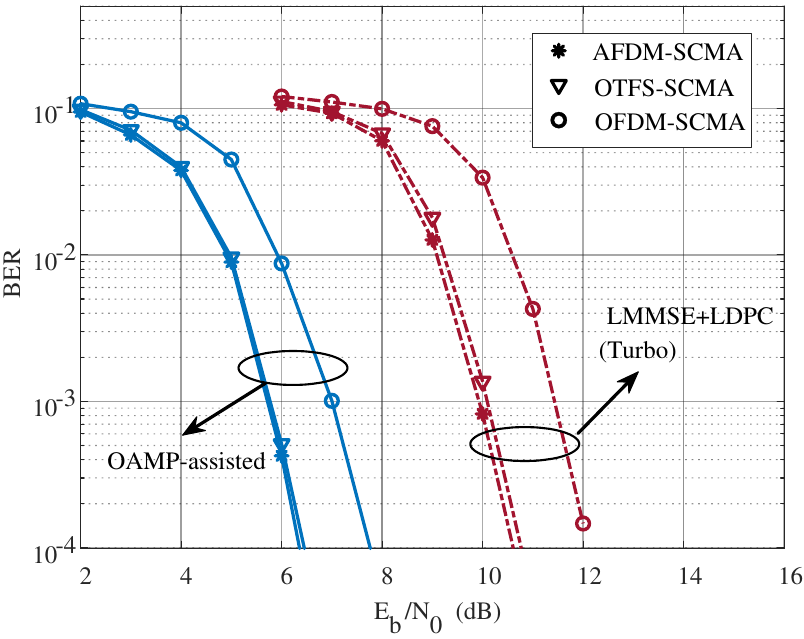}
  \caption{Coded BER   comparison of   AFDM-SCMA,  OTFS-SCMA and OFDM-SCMA   in downlink channels.}\label{OAMP_DL}
  \vspace{-1em}
 \end{figure}

 \begin{figure}[!t]
     \centering
   \includegraphics[width= 0.96\linewidth]{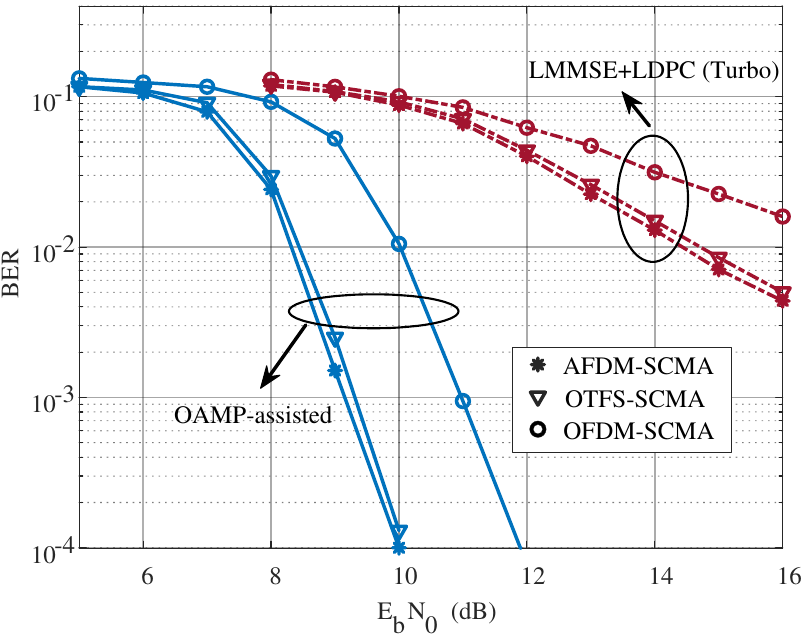}
  \caption{Coded BER   comparison  of   AFDM-SCMA,  OTFS-SCMA and OFDM-SCMA   in uplink channels.}
  \vspace{-1em}
  \label{OAMP_UL}
 \end{figure}

This subsection evaluates  the BER performances of the LDPC-coded  AFDM-SCMA systems over EVA channels. The interleaved transmissions for AFDM-SCMA,  OTFS-SCMA and OFDM-SCMA are considered.   The  5G NR LDPC code, as specified in TS38.212 \cite{5G_NR}, with frame length of $2048$ and rate of $2/3$ is considered.   Table \ref{sim_para}    presents the detailed channel model and system parameters.  The inner LDPC iteration and outer OAMP iteration numbers are set to be $8$ and $10$, respectively. In addition, the damping factor is set as $\kappa = 0.25$.    We also implement the   conventional  turbo receiver   as the benchmark scheme.  The turbo decoding is carried out between  LMMSE and channel decoder by iteratively exchanging soft information in the form of LLR  including \textit{a priori} LLR   and \textit{a posteriori} LLR (extrinsic) \cite{liu2021sparse}. For   fair comparison, the number of iterations in the turbo receiver equals  that of the proposed receiver.

Fig. \ref{OAMP_DL} shows the coded BER performance of the proposed OAMP-enhanced receiver with the proposed I/O inspired-codebook over downlink channels. The two-stage detector proposed in \cite{DekaOTFS_SCMA} is also implemented for comparison.  The main observations are summarized as follows:
\begin{itemize}

    \item Our proposed joint   codebook    and receiver design notably enhance BER performance. Since  MPA   is not required, hence the proposed receiver significantly reduces the detection complexity.  
    \item The proposed receiver significantly outperforms the turbo receiver significantly. For overloaded multiuser systems, the error performance of turbo receiver may degrade significantly since it cannot  guarantee  the orthogonality between the input   and the output errors.

    \item The proposed AFDM-SCMA in conjunction with the proposed receiver  achieves  about   $1.5$ dB  gain over the OFDM-SCMA systems in EVA channels. 

\item  Our proposed  I/O codebook and   OAMP-assisted receiver can  be also employed in OTFS-SCMA for performance enhancement and low-complexity detection.   Notably, the OTFS-SCMA with  the proposed codebook and receiver achieves a similar BER performance with that of the proposed AFDM-SCMA and significantly outperforms OFDM-SCMA.

\end{itemize}

Fig. \ref{OAMP_UL} compares the coded BER performances of the proposed AFDM-SCMA and   OFDM-SCMA   in uplink channels. The advantage of AFDM-SCMA   is more prominent  in the uplink channels, with a noticeable gain of about $2$ dB using the proposed receiver.   Similar to the case of the downlink channels,   the BER performance of the turbo receiver deteriorates significantly.

\begin{figure}[!t]
     \centering
   \includegraphics[width=1\linewidth]{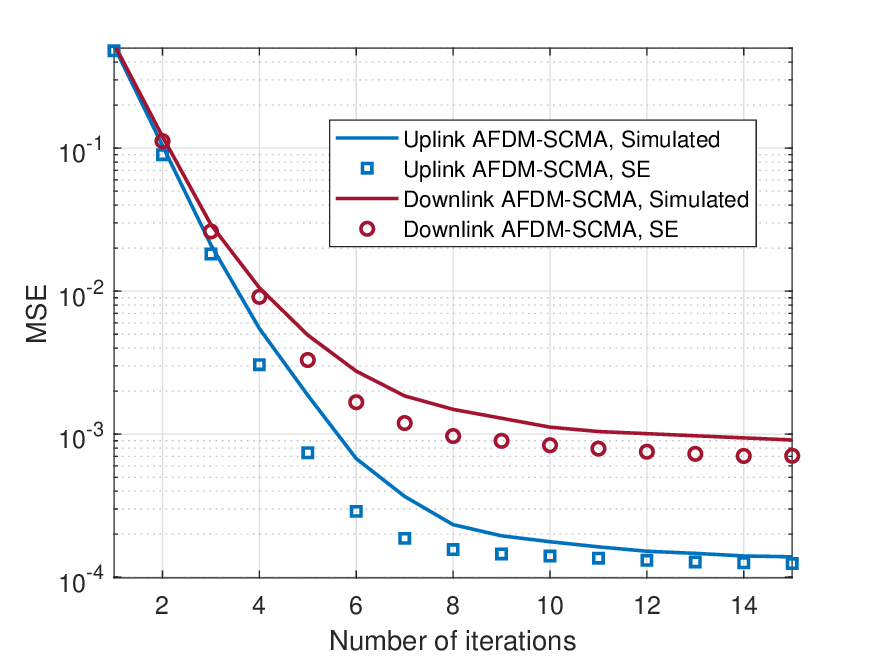}
  \caption{MSE performance v.s. Number of iterations}
  \label{iteration}
  \vspace{-1em}
 \end{figure}

Fig. \ref{iteration} depicts the MSE behaviors with the number of iterations of the proposed receiver in both downlink and uplink channels, where the $E_b/N_0$  values are fixed to be $5$ dB and $9$ dB, respectively.  It is evident that
the simulated MSEs match well  with the predicted MSEs by SE,   demonstrating the effectiveness of SE for analyzing the convergence  that the performance of the proposed
receiver.

\section {Conclusion}
\label{conclu}
 In this paper, we have proposed an AFDM empowered SCMA system, referred to as AFDM-SCMA, for massive connectivity in high mobility channels.  The signal  model, multi-user detection,  system parameters and the BER performance of the proposed AFDM-SCMA have  been discussed in details.  A class of  I/O-inspired codebook   has been proposed, which allows the receiver to utilize  the ZF and LMMSE detectors.  Building upon these codebooks, we have introduced an OAMP assisted  iterative
detection and decoding scheme for both downlink and
uplink channels.   Our numerical results have shown that the proposed AFDM-SCMA   significantly outperforms OFDM-SCMA   in both uncoded and coded systems.   The simulation
results have demonstrated  that the proposed receiver can significantly enhance the BER performance while reducing the decoding complexity.   Moreover,  we have also shown that the proposed receiver can  be utilized in OTFS-SCMA for low-complexity detection.

\ifCLASSOPTIONcaptionsoff
  \newpage
\fi

\bibliography{ref} 
\bibliographystyle{IEEEtran}

\end{document}